\documentclass[usenatbib,useAMS]{mn2e}

\usepackage{natbib}
\usepackage{graphicx,epsfig}
\usepackage{latexsym,amssymb}
\usepackage[fleqn]{amsmath}

\newcommand{\aj}{AJ}%
\newcommand{\apj}{ApJ}%
\newcommand{\apjl}{ApJ}%
\newcommand{\apjs}{ApJS}%
\newcommand{\aap}{A\&A}%
\newcommand{\mnras}{MNRAS}%
\newcommand{\Msun}{M$_\odot$}

\newcommand{\gdm}{\ensuremath{\gamma_\mathrm{dm}}}
\newcommand{\ndm}{\ensuremath{n_\mathrm{dm}}}
\newcommand{\rsdm}{\ensuremath{r_{s,\mathrm{dm}}}}

\newcommand{\Mvdm}{\ensuremath{M_{\mathrm{180},\mathrm{dm}}}}

\newcommand{\cdm}{\ensuremath{c_\mathrm{dm}}}
\newcommand{\badm}{\ensuremath{(b/a)_\mathrm{dm}}}
\newcommand{\cadm}{\ensuremath{(c/a)_\mathrm{dm}}}

\newcommand{\ns}{\ensuremath{n_\star}}

\newcommand{\Mts}{\ensuremath{M_{\mathrm{tot},\star}}}
\newcommand{\Rhs}{\ensuremath{R'_{h,\star}}}

\newcommand{\rv}{\ensuremath{r_\mathrm{180}}}
\newcommand{\Mv}{\ensuremath{M_\mathrm{180}}}

\newcommand{\st}{\ensuremath{\sigma_\mathrm{tot}}}
\newcommand{\avst}{\ensuremath{\langle\sigma_\mathrm{tot}\rangle}}
\newcommand{\stfid}{\ensuremath{\sigma_\mathrm{tot,fid}}}
\newcommand{\sig}[1]{\ensuremath{\sigma_{#1}}}
\newcommand{\avsig}[1]{\ensuremath{\langle\sigma_{#1}\rangle}}
\newcommand{\imsp}{\ensuremath{\Delta\theta}}
\newcommand{\avimsp}{\ensuremath{\langle\Delta\theta\rangle}}
\newcommand{\Rein}{\ensuremath{R_\mathrm{ein}}}

\newcommand{\fpdm}{\ensuremath{f'_\mathrm{dm}}}
\newcommand{\fps}{\ensuremath{f'_{\star}}}
\newcommand{\mathpc}{\mathrm{pc}}
\newcommand{\gravlens}{{\sc gravlens}}

\title[Strong lensing selection biases]{Galaxy density profiles and
  shapes -- II. selection biases in strong lensing surveys}

\author[Mandelbaum, van de Ven, \& Keeton]{ %
  Rachel Mandelbaum$^1$\thanks{\texttt{rmandelb@ias.edu}, Hubble
  Fellow}, %
  Glenn van de Ven$^1$\thanks{\texttt{glenn@ias.edu}, Hubble Fellow}, %
  Charles R.  Keeton$^2$\thanks{\texttt{keeton@physics.rutgers.edu}}
  \\
  $^1$Institute for Advanced Study, Einstein Drive, Princeton NJ
  08540, USA \\ %
  $^2$Department of Physics and Astronomy, Rutgers University, 136
  Frelinghuysen Road, Piscataway NJ, 08854, USA %
}

\date{\today}

\begin{document}

\maketitle 

\begin{abstract}
  Many current and future astronomical surveys will rely on samples of
  strong gravitational lens systems to draw conclusions about galaxy
  mass distributions.  We use a new strong lensing pipeline (presented
  in Paper~I of this series) to explore selection biases that may
  cause the population of strong lensing systems to differ from the
  general galaxy population.  Our focus is on point-source lensing by
  early-type galaxies with two mass components (stellar and dark
  matter) that have a variety of density profiles and shapes motivated
  by observational and theoretical studies of galaxy properties.  We
  seek not only to quantify but also to understand the physics behind
  selection biases related to: galaxy mass, orientation and shape;
  dark matter profile parameters such as inner slope and
  concentration; and adiabatic contraction. We study how all of these
  properties affect the lensing Einstein radius, total cross-section,
  quad/double ratio, and image separation distribution, with a
  flexible treatment of magnification bias to mimic different survey
  strategies.  We present our results for two families of density
  profiles: cusped and deprojected S\'ersic models.  While we use
  fixed lens and source redshifts for most of the analysis, we show
  that the results are applicable to other redshift combinations, and
  we also explore the physics of how our results change for very
  different redshifts.  We find
  significant (factors of several) selection biases with mass;
  orientation, for a given galaxy shape at fixed mass; cusped dark
  matter profile inner slope and concentration; concentration of the
  stellar and dark matter deprojected S\'ersic models.  Interestingly,
  the intrinsic shape of a galaxy does not strongly influence its
  lensing cross-section when we average over viewing angles. Our
    results are an important first step towards understanding how
    strong lens systems relate to the general galaxy population.
\end{abstract}
\maketitle 


\section{Motivation}
\label{S:motivation}

The field of strong lensing is undergoing a period of growth that is
expected to accelerate in the coming decade.  Current samples with
tens of lenses, such as from 
CASTLES\footnote{CASTLES is a collection of uniform HST
  observations of mostly point-source lenses from several samples with
  differing selection criteria, rather than a single,
  uniformly-selected survey.} \citep[e.g.,][]{2001ASPC..237...25F}, 
CLASS \citep[e.g.,][]{2003MNRAS.341...13B},
SDSS \citep[e.g.,][]{2008AJ....135..496I},
and SLACS \citep[e.g.,][]{2008ApJ...682..964B},
will give way to samples with hundreds or even thousands of strong
lensing systems discovered by Pan-STARRS, LSST, SNAP, and SKA
\citep[e.g.,][]{
2004AAS...20510827F,
2004NewAR..48.1085K,
2004ApJ...601..104K,
2005NewAR..49..387M}.
As lens samples grow, they will be even more useful for constraining
the mass distributions of the strong lensing galaxy population
\citep[for a review of strong lensing astrophysics, see][]{Saas-Fee}.

There are, however, certain complications in using strong lensing
studies to constrain the physical properties of galaxies.  The issue
we address here is \emph{selection bias}.  Suppose we treat galaxies
as having an intrinsic probability distribution in some parameter $x$
(such as the inner slope of the density profile), and we want to use
observed strong lens systems to infer that distribution $p(x)$.  If
the lensing probability itself depends on $x$, then in general the
distribution $p_\mathrm{SL}(x)$ for the strong lens galaxies will
not reflect the true, underlying $p(x)$ for all galaxies.  The more
diverse the galaxy population, and the more the strong lensing
cross-section varies across the population, the more important the
selection biases can be.

Obviously, one must account for selection biases in order to draw
reliable conclusions from current and future strong lens samples.  Our
purpose here is to use a complete, end-to-end, and above all, realistic
simulation pipeline for strong lensing (presented in Paper~I of this
series, van de Ven, Mandelbaum, \& Keeton 2008) to quantify and
understand strong lensing selection biases and assess their impact in
typical situations.  We focus on strong lensing of quasars by
early-type, central (non-satellite) galaxies at several mass scales,
using two-component mass profiles (dark matter plus stars) that are
consistent with existing photometry and stacked weak lensing data from
SDSS \citep{2003MNRAS.341...33K,2006MNRAS.368..715M} as well as with 
$N$-body and hydrodynamic simulations.

The issue of selection biases in strong lensing is not a new one,
and we are building on many previous studies of this issue.  The
main galaxy properties that we study in an attempt to understand
selection bias are
\begin{itemize}
\item Mass \citep[e.g., ][]{1984ApJ...284....1T, 1991MNRAS.253...99F,
    2007MNRAS.379.1195M};
\item Orientation \citep[``inclination bias'', e.g.,
  ][]{1997ApJ...486..681M, 1998ApJ...495..157K, 2007arXiv0710.1683R};
\item Shape \citep[e.g., ][]{1997ApJ...482..604K, 2005ApJ...624...34H, 2007arXiv0710.1683R};
\item Dark matter inner slope 
  \citep[e.g., ][]{2001ApJ...549L..25K, 2001ApJ...555..504W,2002ApJ...566..652L};  
\item Dark matter concentration  
  \citep[e.g., ][]{2004ApJ...601..104K, 2007A&A...473..715F}.
\end{itemize}
In most cases, previous studies addressed the issue of selection bias
using models that were simplified in some way. Simplifications often
included testing effects of dark matter slope and concentration using
pure (generalised) NFW profiles without a baryonic component; or testing
effects of density profile or shape using single-component mass
models.  A notably different approach was taken by
\citet{2007MNRAS.382..121H}, \citet{2007MNRAS.379.1195M}, and
\citet{2008MNRAS.386.1845H}, who used the 
Millennium simulation with a 
semi-analytic model of galaxy formation to simulate  strong
lensing.  That approach naturally yields a realistic
distribution of halo and galaxy properties (to the extent that the
true distribution of galaxy properties and their relation to halo
properties is encoded in the semi-analytic
model), at least for a fixed cosmological 
model, and it will undoubtedly be useful for modeling the strong
lensing population in large, future surveys.  We elect, however, to
take a more controlled approach here, using realistic but discrete
values of the density profile parameters, so that we can disentangle
the different types of lensing selection biases, and understand the
physics of each one, before recombining them to determine the net
effects.

Our basic approach is to generate realistic galaxy models with various
values of the relevant parameters, and to investigate how the lensing
cross-section $\sigma(\vec{x})$ depends on model parameters
  $\vec{x}$ (where $\vec{x}$ might include mass, shape, concentration,
  and other parameters, some of which may be correlated). The reason
for focusing on the cross-section is that it represents the weighting
factor that transforms the intrinsic joint parameter distribution
$p(\vec{x})$ into the distribution $p_\mathrm{SL}(\vec{x})$ among
observed strong lens systems. Note that we focus on selection biases
related to \emph{physical} effects.  There may also be
\emph{observational} selection biases \citep[e.g.,
][]{1991ApJ...379..517K}, but they are specific to a given survey and
are not something that we can address in a general way.   This
  fact is one reason for our focus on point-source lensing: extended
  source lensing is even more sensitive to observational selection
  effects than quasar lensing. For example, the effect of a finite
  aperture is no longer straightforward for an extended source, and
  detection of an extended strong lens system depends on factors such
  as the size of the point-spread function which varies spatially and
  temporally in any given survey.  Fortunately, with large numbers of
  point-source lens systems anticipated in future surveys [e.g., $\sim
  10^{5}$ with SKA \citep{2004NewAR..48.1085K}] this investigation of
  physical selection biases in point-source lens systems should be
  highly useful. We emphasize that our purpose is only to determine
  the mapping function $\sigma(\vec{x})$ due to physical differences
  between galaxies, and not to predict the observed distribution of
  properties $p_\mathrm{SL}(\vec{x})$ in some particular lensing
  survey, for which knowledge of the intrinsic parameter distribution
  $p(\vec{x})$ and of any observational selection effects are both
  necessary.

In this paper, we present a systematic investigation of
physical strong lensing selection biases that both unifies and extends
previous work.  We begin in Section~\ref{S:simulations} with a brief review
of the simulation pipeline developed in Paper~I, including the density
profiles and shapes of our galaxy models, our computational lensing
methods, and the basic lensing properties of our galaxies.  In
Section~\ref{S:orientationshape}, we investigate selection biases related
to galaxy orientation and shape.  In Section~\ref{S:innerslope}, we consider
selection biases related to the inner slope of the dark matter
component of the density profile (for cusped models).  In
Section~\ref{S:concentration}, we allow the dark matter concentration to
vary as well.  In Section~\ref{S:sersic}, we turn to deprojected S\'ersic
density profiles and examine selection biases with both the stellar
and dark matter parameters. Section~\ref{S:robustness} includes an exploration of the 
ranges of lens and source redshifts for which our conclusions are
applicable.  Finally, in Section~\ref{S:conclusions}, we summarise our main
conclusions and discuss their implications for past and future strong
lensing analyses.

\section{Simulations}
\label{S:simulations}

Here we summarise the strong lensing simulation pipeline that was
developed in Paper~I. For more details regarding the choice of galaxy
models, and the lensing calculations (including validation and
convergence tests), we refer the reader to that paper.

\subsection{Notation and conventions}\label{SS:notation}

The pipeline assumes fiducial lens and source redshifts of $z_L=0.3$
and $z_S=2.0$, respectively, although we explore a range of lens
and source redshifts in Section~\ref{S:robustness}.
We compute distances using a flat $\Lambda$CDM cosmology with
$\Omega_m=0.27$ and $h=0.72$.  With these choices, the angular
scale at the lens redshift is $1\arcsec = 4.36$ kpc and the
lensing critical density is $\Sigma_c=2389\,M_\odot\,\mathpc^{-2}$.

We use $x$, $y$, and $z$ to denote intrinsic, three-dimensional (3d)
coordinates, and $x'$ and $y'$ for the projected, two-dimensional (2d)
coordinates. Similarly, $r$ is the intrinsic radius
($r^2=x^2+y^2+z^2$) and $R'$ is the projected radius
($R'^2=x'^2+y'^2$). For non-spherical galaxies, we use $a$, $b$, and
$c$ for the major, intermediate, and minor semi-axis lengths of the
intrinsic, 3d density profiles; and we use $a'$ and $b'$ for the major
and minor semi-axis lengths of the projected, 2d surface densities.
Subscripts ``dm'' and ``$\star$'' are used to indicate whether a
quantity describes the dark matter (DM) or stellar component of the
galaxy model.  Any gas component that has not cooled to form stars ---
which is expected to be sub-dominant relative to the stellar component
for early type galaxies \citep[e.g., ][]{2005RSPTA.363.2693R,
  2006MNRAS.371..157M} --- is implicitly included in the component
that we label dark matter.

All quoted dark matter masses $\Mv$ are defined using comoving
quantities such that the average density within the virial radius is
$180\overline{\rho}$.  This definition is in principle arbitrary, but
this choice was used to analyse the weak lensing results that we use
to estimate masses for our galaxy models.

\subsection{Galaxy models}\label{SS:galaxymodels}

Our guiding principle in designing the galaxy models and choosing
  their parameters is that we want them to be realistic examples of
  galaxies, but not necessarily to reproduce the full intrinsic parameter
  distribution of real galaxies.  The reason for the latter point is
  that strong lensing is inherently non-linear, so it may be the case
  that a region of parameter space with very few galaxies is actually
  quite important for strong lensing because it has a large
  cross-section. A classic example is brightest cluster galaxies,
  which are a negligible fraction of the galaxy population, yet
  constitute a non-negligible fraction of strong lenses.
  Consequently, we choose parameter values taking into account prior
  knowledge of the intrinsic parameter distributions ---where known---
  and using physical intuition for what effects may strongly increase
  the lensing cross-section, such as a steep inner slope of the dark
  matter density.

\subsubsection{Basic setup}

All of the ellipsoidal galaxy models used for this pipeline involve two
components, one for the stars and one for the dark matter.  The
parameters of stellar components are based on photometry and
spectroscopy from the SDSS \citep{2003MNRAS.341...33K}.  The
parameters of the dark matter components are derived from $N$-body
simulations and stacked weak lensing observations of early-type
galaxies in SDSS \citep{2006MNRAS.368..715M}.
We attempt to approximately bracket the range of luminosities of
observed strong lensing systems by using models corresponding to $\sim
2L_*$ and $\sim 7L_*$ early-type, central (non-satellite) lens
galaxies.  (See, e.g., \citealt{2004ApJ...612..660K} and
\citealt{2006ApJ...641..169M} for a discussion of environmental
effects when the lens galaxy is a satellite in a group or cluster of
galaxies.) The derivation of the resulting mass and length scales is
explained in Paper~I. 

Here we briefly mention that the stellar components have masses $\Mts$
of $1.16$ and $5.6\times 10^{11}$\,\Msun, and projected half-mass
radii $\Rhs$ of $4.07$ and $11.7$\,kpc for the above lower
(``galaxy'') and higher (``group'') mass scale, respectively.  The
dark matter components have halo masses $\Mvdm$ of $0.47$ and
$9.3\times 10^{13}$\,\Msun, and concentrations $\cdm$ of $8.4$ and
$5.6$, which implies scale radii $\rsdm$ of $49.7$ and $201$\,kpc, for
the lower and higher mass scale, respectively.   We note that a
  simulation with only two mass models is not sufficient for a fully
  quantitative study of mass bias in strong lensing. We have limited
  the number of mass models because mass bias is among the oldest and
  best-studied selection effects in strong lensing, so we focus
  instead on other effects for which the answers are less clear.

\subsubsection{Density profiles}

We use two basic types of density profiles for the simulation
pipeline. The first are cusped density profiles, which have the
general form
\begin{equation}
  \label{eq:denscusp}
  \rho(r) = \frac{\rho_0}{m^\gamma (1+m)^{n-\gamma}},
\end{equation}
where $m$ is a dimensionless ellipsoidal radius; in the spherical
case, $m=r/r_s$ for some scale radius $r_s$.  For the stellar cusped
density, we use the \citet{1990ApJ...356..359H} profile,
which has $(\gamma,n)=(1,4)$.  For the cusped density of the DM halo,
we use the NFW \citep{1997ApJ...490..493N} profile, which has
$(\gamma,n)=(1,3)$, but generalise it by allowing $\gamma$ to vary on
a grid of values $\{0.5,1,1.5\}$. Further variation of the cusped
models includes several simulations with adiabatic contraction
of the DM halo due to the presence of the stellar component, according
to the prescriptions given in \cite{1986ApJ...301...27B} and
\cite{2004ApJ...616...16G}.

The second type of profile that we use is a deprojected
\citet{1968adga.book.....S} profile, which is well-described using the
formula from \cite{1997A&A...321..111P},
\begin{eqnarray}
  \label{eq:rhosersic}
  \rho(m) & = & \frac{\rho_0}{m^{p_n}} \; 
  \exp\left[-b_n\,m^{1/n} \right] ,
\end{eqnarray}
where the central (negative) logarithmic slope is related to the
S\'ersic index $n$ by $p_n = 1 - 0.6097/n + 0.05563/n^2$, and $b_n =
2n - 1/3 + 4/(405n) + 46/(25515n^2)$ to high precision
\citep{1999A&A...352..447C}. The scale radius is the projected
half-mass radius, which in case of a constant stellar mass-to-light
ratio corresponds to the effective radius $R_e$. The S\'ersic indices
$\ns$ of the stellar component from observational constraints
\citep{2006ApJS..164..334F} and $\ndm$ of the dark matter component
from $N$-body simulations \citep{2005ApJ...624L..85M} are $4.29$ and
$2.96$ (respectively) for the lower mass scale, and $5.87$ and $2.65$
for the higher mass scale.  We also investigate varying both $\ndm$
and $\ns$ away from these fiducial values within the ranges derived in
Paper~I.

Paper~I includes a number of figures characterising the intrinsic (3d)
and projected (2d) density profiles of our model galaxies.
We highlight the following projected quantities (since the
projected mass distribution is what governs lensing): the surface
mass density $\Sigma(R')$ (figure 4); the negative logarithmic slope
$\gamma'(R')$ of the surface mass density (figure 5); the projected
enclosed dark matter fraction $\fpdm(R')$ (figure 6); and the lensing
deflection angle $\alpha(R')$ (figure 7).  Those quantities are
plotted for both the cusped and deprojected S\'ersic density profiles
at both mass scales.

One characteristic of these galaxy models that was noted in
  Paper~I, and that will be important in our interpretation of results
  in this paper, is that from $\sim 1$--$50$ kpc (cusped models) or
  from $\sim 0.3$--$30$ kpc (deprojected S\'ersic models), the
  composite galaxy model is approximately isothermal. This result can
  be most easily seen in Paper~I figures 5 and 7. 
  Moreover, it is still nearly true not only for our fiducial models
  but for our models with varying $\gdm$ (cusped) or $\ndm$ and $\ns$
  (S\'ersic). For the higher mass models, 
the deviations from isothermality are more significant. 

  While we will find later on that the deviations from isothermality
  on small scales, well within $\Rein$, can have noticable effects for
  the lensing properties (e.g., the relationship between the Einstein
  radius and the total unbiased cross section differs from the
  relation $\sigma = \pi \Rein^2$ that applies to singular isothermal
  spheres), we can nonetheless use this approximate result to reason
  about selection biases with certain galaxy properties. For example,
  steeper dark matter inner slopes $\gdm$ and higher concentrations
  $\cdm$, $\ndm$, or $\ns$ mean that a larger fraction of the total
  galaxy mass (stellar plus dark) is located within some fixed
  aperture. When combining this fact with the approximate
  isothermality, we can envision the profile on those scales as simply
  being increased by a multiplicative factor when we increase $\gdm$,
  $\cdm$, $\ndm$, or $\ns$. Consequently, higher values of these
  parameters increase the total lensing cross-section, and in the case
  of $\gdm$, $\cdm$, and $\ndm$, they also increase the projected dark
  matter fraction $\fpdm$ in the inner regions.

\subsubsection{Shape models}

We consider seven different models for the shape of the galaxy
density: spherical, moderately oblate, very oblate, moderately
prolate, very prolate, triaxial, and a mixed model. In all but the
mixed model, the axis ratios used for the dark matter and stellar
components are related in a simple way (with the stellar component
being slightly rounder than the dark matter component). In all cases
the axes of the two components are intrinsically aligned. The mixed
model includes an oblate stellar component that has short axis aligned
with the short axis of a triaxial dark matter halo.  The chosen axis
ratios for all models, and the method of sampling in viewing angle to
compute accurate orientation-averaged lensing cross-sections, are
given in Paper~I.

\subsection{Strong lensing calculations}\label{SS:stronglensing}

Here we briefly summarise the strong lensing calculations, which are
described in far greater detail in Paper~I.  In all cases, we consider
only point-source (quasar) lensing, and do all calculations using an
updated version of the \gravlens\ software package
\citep{2001astro.ph..2340K}.

For the cusped models, we compute the scaled surface mass density
$\kappa = \Sigma/\Sigma_c$ on a map, and use Fourier transforms to
solve the Poisson equation and find the lens potential.  We then
compute the first and second real-space derivatives of the potential
in Fourier space, and use inverse FFTs to transform back to real
space.  Paper~I presents tests to determine what map resolution
and size are necessary to avoid numerical errors in the Fourier
analysis.  With deprojected S\'ersic models it is hard for the
map-based approach to resolve the steep central profile, so we instead
compute the lensing deflection and magnification analytically
\citep[for spherical models; also see][]{2004A&A...415..839C,
2009JCAP...01..015B}, or with numerical integrals (for non-spherical
models).

We use the routine \texttt{calcRein} to calculate the Einstein radius
$\Rein$ for general density distributions, and the routine
\texttt{mockcsec} to compute strong lensing cross-sections using Monte
Carlo techniques.  Specifically, \texttt{mockcsec} places many random
sources behind the lens, solves the lens equation to find all the
images, and then tabulates the cross-sections for lensing with
different numbers of images (2, 3, or 4).\footnote{In each 2-image and
  4-image system, there is actually one additional image that lies
  near the centre of the lens galaxy (so in some of the lensing
  literature such systems are called 3-image and 5-image systems
  instead).  The central images are highly de-magnified and rarely
  observed, however, which is why most observed lenses are considered
  to be doubles and quads.  We do find the central images but do not
  include them in our astrophysical analysis.  See Paper I for more
  discussion.}  We can add those cross-sections to find the total
lensing cross-section, \st.  We use bootstrap resampling to estimate
the statistical uncertainties in the cross-sections.

We account for magnification bias in various ways intended to mimic
different survey strategies.  We use the \gravlens\ routine
\texttt{mockLF} to define an observationally-motivated
double-power-law luminosity function for the source quasar population;
and we consider surveys with faint, intermediate and high flux limits
(ranging from locally shallow to locally steep parts of the luminosity
function: $0.04$, $0.4$, and $4L_*$).  We compute magnification bias using either the total
magnification of all images, which is appropriate for surveys that do
not initially resolve strong lens systems, and separately, using the magnification of
the second-brightest image (for surveys with sufficient resolution to
resolve strong lens systems from the outset).  All told, we examine
six types of biased cross-sections in addition to the unbiased lensing
cross-section.

\subsection{Basic results}\label{SS:basicresults}

Here, we present the basic lensing properties of our spherical
cusped galaxy models.  Table~\ref{T:basic} gives the Einstein radii
and unbiased lensing cross-sections ($\st=\sig{2}$ for spherical
models) for models at both mass scales, for different values of the
dark matter inner slope $\gdm$.

\begin{table*}
\caption{\label{T:basic}Summary of the basic lensing results for the spherical
  shape model.}
\begin{tabular}{l|cccc|cccc}
\hline
\hline
Description & \multicolumn{4}{c}{Lower mass} &
\multicolumn{4}{c}{Higher mass} \\
 & $\gdm=0.5$ & $\gdm=1$ & $\gdm=1.5$ & SIS & $\gdm=0.5$ & $\gdm=1$ &
 $\gdm=1.5$ & SIS \\
\hline
$\Rein$ (kpc) & $2.5$ & $3.0$ & $4.5$ & $2.4$ & $5.1$ & $7.6$ & $19.4$
& $17.3$ \\
$\Rein$ (arcsec) & $0.58$ & $0.68$ & $1.04$ & $0.55$ & $1.18$ & $1.75$
& $4.44$ & $3.97$ \\
$\st$ (arcsec$^2$), unbiased & $0.42$ & $0.53$ & $1.40$ & $0.95$ & $0.84$ &
$1.63$ & $15.0$ & $49.5$ \\
\hline
\end{tabular}
\end{table*}

There are three points to note about these results.  First, the strong
lensing properties are, as expected, a strong function of mass.  For
the fiducial dark matter inner slope $\gdm=1$, the Einstein radius
$\Rein$ and lensing cross-section increase by factors of $2.6$ and
$3.1$ when we go from the lower to the higher mass model; for
$\gdm=1.5$, the factors of $4.3$ and $10.7$ increase are even larger.
Our results are broadly consistent with conclusions by
\cite{1984ApJ...284....1T}, \citet{1991MNRAS.253...99F},
\citet{2007MNRAS.379.1195M} and from many other studies that the
population of lens galaxies differs from the overall galaxy population
in its overall higher mean stellar and DM halo mass. The higher mean
mass for the strong lensing systems is determined by the competition
between the rising $\st$ with mass and the steeply declining mass
function at high masses.  A second point is that the increase in $\st$
with mass is notably larger for the models with $\gdm=1.5$ than for
the models with $\gdm=1$.  This means the strong lensing mass bias
depends sensitively on the inner slope of the intrinsic density
profile.  A third point is that, even when the mass is fixed, $\Rein$
and $\st$ are very sensitive to $\gdm$.  While this is not surprising,
the latter two points do serve as a reminder that calling strong lens
galaxies a ``mass-selected'' sample of galaxies is an over-simplification.

As a reference point, we also consider a single-component singular
isothermal sphere (SIS) model which has the same mass within the
virial radius as our two-component galaxy models (for each mass
scale).  Even if the density profile is truncated at the virial
radius, for radii well inside the virial radius the intrinsic density
profile can be given as $\rho_{SIS}(r) = \Mv /(4\pi \rv r^2)$, and the
surface density as $\Sigma(R') \simeq \Mv/(4\rv R')$.  To find the
Einstein radius, we require for the mean surface brightness that
$\overline{\Sigma}(\Rein) = \Sigma_c$, which gives
\begin{equation}
  \Rein^{(\mathrm{SIS})} = \frac{\Mv}{2\rv\Sigma_c} \propto \Mv^{2/3}
\end{equation}
where $\Mv$ and $\rv$ are related by our virial mass definition,
\begin{equation}
  \frac{3\Mv}{4\pi \rv^3} = 180\overline{\rho}
\end{equation}
The resulting Einstein radii and lensing cross-sections
($\st = \pi \Rein^2$) are also given in Table~\ref{T:basic}.

Comparing the familiar SIS model with our more complex, two-component
models yields some notable results.  For the lower mass scale, the SIS
Einstein radius is similar to that of our shallow $\gdm=0.5$ cusped
model, but the SIS cross-section lies between those of the steeper
$\gdm=1$ and $\gdm=1.5$ cusped models.  For the higher mass scale, the
SIS Einstein radius $\Rein^{(\mathrm{SIS})}$ is only slightly smaller
than that for the $\gdm=1.5$ cusped model, and the SIS cross-section
is considerably higher than that for any of our cusped models.  This
result can be understood from  paper~I figure 5, 
which shows
that while $\gamma'$ for our lower mass models is near the SIS value of
$\gamma'=1$ near $\Rein$, it is significantly shallower for a
fair fraction of the range inside $\Rein$. The shallower inner
slope moves the (source-plane) radial caustic inwards, decreasing the
cross-section to well below the SIS value of $\pi \Rein^2$.

One implication is that a simple SIS model cannot mimic any particular
one of our cusped models.  Thus, even though galaxies seem to have
quasi-isothermal profiles in the vicinity of the Einstein radius (see
Paper~I), a simple SIS is still not very realistic as a \emph{global}
lens model.  The difference between the SIS and more complex two-component
lens models is particularly notable on mass scales corresponding to
groups of galaxies and above \citep[also see][]{1998PhDT.........6K,
  2000ApJ...532..679P,2001ApJ...559..531K,2004ApJ...601..104K}.

\section{Orientation and shape}
\label{S:orientationshape}

We begin our main science results by examining selection biases as a
function of galaxy orientation and shape, for fixed dark matter inner
slope \gdm.  In Section~\ref{SS:resultsorient} we investigate ``orientation
bias'' for each of the six non-spherical shape models.  In
Section~\ref{SS:resultsshape} we consider orientation-averaged
cross-sections for all model shapes, to see if there are significant
selection biases associated with the \emph{intrinsic} shapes of
galaxies.  Finally, in Section~\ref{SS:resultsmagbias} we consider whether
there is any simple, universal conversion between unbiased and biased
lensing cross-sections.

\subsection{Orientation}
\label{SS:resultsorient}

\begin{figure}
\begin{center}
  \includegraphics[width=\columnwidth,angle=0]{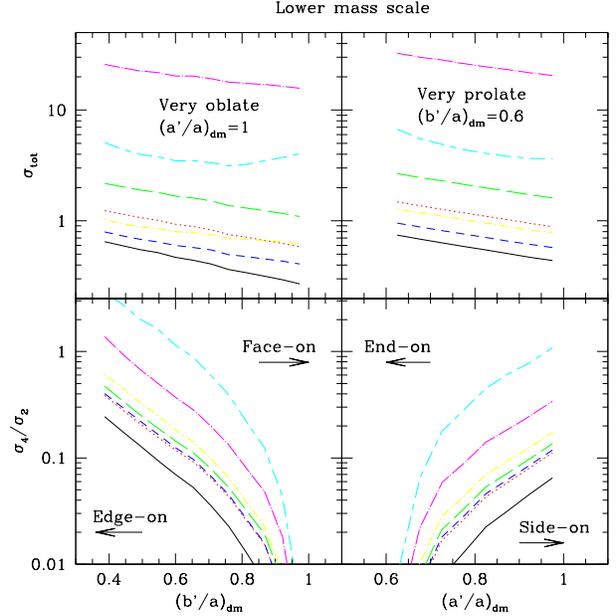}
\caption{\label{F:incoblpro}
  The total lensing cross-section \st\ (top) and quad/double ratio
  (bottom) as a function of projected semi-axis lengths for the lower
  mass cusped model for very oblate (left) and very prolate (right)
  shapes.  The lines indicate magnification bias modes as follows:
  black solid $=$ unbiased, red dotted and blue dashed $=$ weighted by
  total and second-brightest magnification respectively for $0.04L_*$
  limiting magnitude, green long-dashed and yellow dot-dashed $=$ the
  same for $0.4L_*$ limiting magnitude, and magenta dot-long dashed
  and cyan short-long dashed $=$ the same for $4L_*$ limiting
  magnitude.}
\end{center}
\end{figure}
\begin{figure}
\begin{center}
  \includegraphics[width=\columnwidth,angle=0]{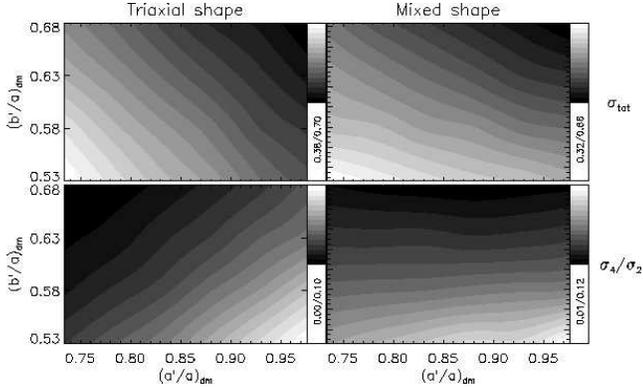}
\caption{\label{F:inctriaxmix}
  Contour plots of the total lensing cross-sections $\st$ (top) and quad/double
  ratios (bottom) as a function of projected semi-axis lengths for
  the lower mass cusped model for triaxial (left) and mixed (right)
  shapes, where the axis labels indicate the shape of the dark matter
  component.  All quantities are unbiased cross-sections only.}
\end{center}
\end{figure}

In this subsection, we fix the dark matter inner slope to $\gdm=1$
and examine how the orientation of non-spherical models affects their
lensing cross-sections (``orientation bias'').  Our basic physical
intuition suggests the following trends with orientation for a given
shape:
\begin{enumerate}
\item When viewing the galaxy along the long axis (with scale length
  $a$), the projected scale radii are minimised, so that the most mass
  is squeezed into the smallest projected area. The resulting higher
  surface density at fixed $R'$ implies an increased Einstein radius
  and higher lensing cross-section.
\item The four-image cross-section $\sig{4}$ and the quad/double ratio
  $\sig{4}/\sig{2}$ increase with the projected ellipticity; they are
  maximised when viewing the halo along the intermediate axis, such
  that the long and short axes are seen in projection (i.e., projected
  axis ratio $b'/a'=c/a$).  The four-image cross-section vanishes when
  the projected surface density has circular symmetry, i.e., for an
  oblate galaxy viewed face-on or a prolate galaxy viewed end-on.
\item The three-image cross-sections can only be significant in cases of
  extreme projected ellipticity, when the tangential caustic protrudes
  outside the radial caustic.
\end{enumerate}

We first consider the very oblate and very prolate models.
Fig.~\ref{F:incoblpro} shows the total cross-section $\st$ and the
quad/double ratio $\sig{4}/\sig{2}$ for the lower mass scale (results
for the higher mass scale are similar).  The total cross-section is
indeed higher for oblate models viewed edge-on rather than face-on,
and for prolate models viewed end-on rather than side-on (i.e., viewed
with the long axis along the line-of-sight in each case).  The total
cross-section varies with orientation by factors of $\sim 2$ for these
strongly non-spherical models; in the higher mass model (not shown),
the variation is as much as a factor of $4$ because the flatter DM
component is more prominent.  This finding implies that if strong
lensing systems seen in reality are drawn from significantly
non-spherical galaxies, we should observe preferential alignment of
the long axis with the line-of-sight, rather than random orientations.

The cross-sections for three- and four-image systems behave in
accordance with our expectations. As can be seen from
$\sig{4}/\sig{2}$ in the bottom panels of Fig.~\ref{F:incoblpro},
four-image systems are plentiful for orientations that yield a large
projected ellipticity --- for oblate galaxies viewed edge-on, or
prolate galaxies viewed side-on --- but they naturally disappear as
the projected mass density approaches circularity.  While we have not
plotted $\sig{3}$ since in most cases it is zero, we see substantial
numbers of three-image systems only for edge-on oblate galaxies, where
the projected matter density is so flat that the tangential caustic
protrudes beyond the radial caustic. Above some critical projected
axis ratio, $\sig{3}$ becomes consistent with zero.

Magnification bias increases the effective cross-sections, because
lensing magnification brings in systems that would have been below the
flux limit had they not been lensed.  The effect is strongest when the
flux limit is in the steep part of the source luminosity function
(the $4L_*$ limiting luminosity).  In all cases, the biased
cross-section is larger when calculated using the total magnification
rather than the magnification of the second-brightest image, but the
trends with viewing angle are typically preserved.  The
$\sig{4}/\sig{2}$ ratio is increased by magnification bias, because
four-image lenses tend to have higher magnifications than two-image
lenses.  This ratio increases with the flux limit, but it is actually
higher when using the second-brightest magnification rather than the
total magnification, because many four-image lenses have at least two
bright images.  We discuss further details of the biased
cross-sections in Section~\ref{SS:resultsmagbias}.

Next, we consider the more complex triaxial and mixed shape models
shown in Fig.~\ref{F:inctriaxmix}. For both mass scales, the total
cross-sections have maxima when $a'/a = b/a$ and $b'/a = c/a$ (i.e.,
the projected semi-axis lengths have their minimum values), which
corresponds to the long axis being along the line-of-sight. This result is
therefore physically intuitive, and consistent with the results in
\cite{2007arXiv0710.1683R} for single-component isothermal ellipsoid
lens models. The total variation of \st\ with viewing angle for
these typical triaxial models is a factor of $\sim 2$ at the lower
mass scale, or as much as $4$ in the higher mass case (where the
flatter DM component contributes more).  The $\sig{4}/\sig{2}$
ratio reaches its maximum value when $a'/a=1$ (maximum value)
and $b'/a = c/a$ (minimum value), which corresponds to the most
flattened projected mass distribution. In the mixed case, the
orientation of the oblate stellar component is such that, regardless
of $(a'/a)_\mathrm{dm}$, the top of the plot corresponds to a
rounder projected stellar density because it is seen face-on,
while the bottom of the plot corresponds to a flatter projected
stellar density because it is seen edge-on, so that both $\st$
and $\sig{4}/\sig{2}$ increase from top to bottom. As a result,
the trends in both $\st$ and $\sig{4}/\sig{2}$ are not significantly
modified from the triaxial case, except for a tilt in the contours
which can be accounted for by considering the orientation of the
oblate stellar component.

We conclude that strong lenses drawn from a parent population with
randomly-oriented non-spherical matter distributions have a projected
axis ratio distribution that is not consistent with that of a random
sample of the population, because there are preferred orientations for
the lensing systems.  The degree of preference for particular
orientations can be factors of several depending on the intrinsic axis
ratios, the relative contribution of the flatter DM component, and the
number of images in the lenses being considered. While it is difficult
to compare quantitatively with the conclusions of previous works on
``inclination bias''
\citep[e.g.,][]{1997ApJ...486..681M,1998ApJ...495..157K,
  2007arXiv0710.1683R} due to the very different mass models used,
qualitatively we agree with the results that for an intrinsically
non-spherical model, orientation effects are important at this level.
In particular, the trends we find regarding $\st$ and
$\sig{4}/\sig{2}$ for triaxial models agree well with those of
\cite{2007arXiv0710.1683R}.

The preceding discussion has used results for fixed dark matter inner
slope $\gdm=1$. Generally, we find that the orientation bias trends
in $\sig{4}$ are somewhat weaker for shallower $\gdm=0.5$ models, and
stronger for steeper $\gdm=1.5$ models, at the level of 10 to 30 per
cent; whereas for $\sig{2}$ the degree of orientation bias is roughly
the same for $\gdm=0.5$ and for $\gdm=1.5$.  This may occur because
using a shallower \gdm\ makes the rounder stellar component more
significant for the lensing properties, decreasing the overall
flattening and hence decreasing the four-image cross-section.

\subsection{Shape}
\label{SS:resultsshape}

Next we study \emph{orientation-averaged} lensing cross-sections to
see whether there are significant biases related to the intrinsic
shape of the galaxy.  Unless explicitly noted, we again use our
fiducial cusped models with $\gdm=1$.  Table~\ref{T:haloshape} lists
the total cross-sections and quad/double ratios of all seven shape
models, for each magnification bias mode and for both mass scales.
(Note that in some cases, the statistical errors are not identically
zero but round to zero at the precision quoted in the table.)

\begin{table*}
\begin{center}
\caption{Lensing cross-sections averaged over all orientations, for each model shape and
  magnification bias mode, for the fiducial dark matter concentration
  and inner slope $\gdm = 1$, and for both mass scales. \label{T:haloshape}}
\begin{tabular}{llrrrr}
\hline\hline
 & & \multicolumn{2}{c}{Lower mass} & \multicolumn{2}{c}{Higher mass} \\
Shape & Bias mode & $\avst$ (arcsec$^2$) & $\avsig{4}/\avsig{2}$ & $\avst$ (arcsec$^2$) & $\avsig{4}/\avsig{2}$ \\ 
\hline
Spherical & Unbiased & $ 0.53 \pm  0.00$ & $ 0.000 \pm  0.000$ & $ 1.63 \pm  0.00$ & $ 0.000 \pm  0.000$\\
 & Faintest, total & $ 1.08 \pm  0.00$ & $ 0.000 \pm  0.000$ & $ 4.18 \pm  0.01$ & $ 0.000 \pm  0.000$ \\
 & Faintest, second & $ 0.72 \pm  0.00$ & $ 0.000 \pm  0.000$ & $ 2.97 \pm  0.01$ & $ 0.000 \pm  0.000$ \\
 & Medium, total & $ 1.97 \pm  0.01$ & $ 0.000 \pm  0.000$ & $ 8.34 \pm  0.02$ & $ 0.000 \pm  0.000$ \\
 & Medium, second & $ 1.02 \pm  0.01$ & $ 0.000 \pm  0.000$ & $ 5.09 \pm  0.02$ & $ 0.000 \pm  0.000$\\
 & Brightest, total & $25.12 \pm  0.22$ & $ 0.000 \pm  0.000$ & $163.52 \pm  0.85$ & $ 0.000 \pm  0.000$ \\
 & Brightest, second & $ 5.78 \pm  0.10$ & $ 0.000 \pm  0.000$  & $51.56 \pm  0.56$ & $ 0.000 \pm  0.000$ \\
\hline
Moderately oblate & Unbiased & $ 0.49 \pm  0.00$ & $ 0.054 \pm  0.001$ & $ 1.46 \pm  0.01$ & $ 0.264 \pm  0.004$ \\
 & Faintest, total & $ 0.98 \pm  0.00$ & $ 0.089 \pm  0.001$ & $ 3.65 \pm  0.02$ & $ 0.366 \pm  0.005$ \\
 & Faintest, second & $ 0.64 \pm  0.00$ & $ 0.093 \pm  0.001$ & $ 2.47 \pm  0.01$ & $ 0.362 \pm  0.005$ \\
 & Medium, total & $ 1.77 \pm  0.01$ & $ 0.107 \pm  0.002$ & $ 7.21 \pm  0.04$ & $ 0.406 \pm  0.006$ \\
 & Medium, second & $ 0.87 \pm  0.00$ & $ 0.132 \pm  0.002$ & $ 4.08 \pm  0.02$ & $ 0.429 \pm  0.006$ \\
 & Brightest, total & $22.09 \pm  0.09$ & $ 0.251 \pm  0.004$ & $138.28 \pm  0.78$ & $ 0.645 \pm  0.009$ \\
 & Brightest, second & $ 4.01 \pm  0.02$ & $ 0.611 \pm  0.009$ & $36.16 \pm  0.23$ & $ 1.055 \pm  0.013$ \\
\hline
Very oblate & Unbiased & $ 0.47 \pm  0.00$ & $ 0.110 \pm  0.002$ & $ 1.48 \pm  0.01$ & $ 0.540 \pm  0.010$ \\
 & Faintest, total & $ 0.93 \pm  0.00$ & $ 0.168 \pm  0.002$ & $ 3.73 \pm  0.03$ & $ 0.684 \pm  0.012$ \\
 & Faintest, second & $ 0.60 \pm  0.00$ & $ 0.173 \pm  0.002$ & $ 2.51 \pm  0.02$ & $ 0.662 \pm  0.011$ \\
 & Medium, total & $ 1.66 \pm  0.01$ & $ 0.198 \pm  0.003$ & $ 7.38 \pm  0.06$ & $ 0.742 \pm  0.013$ \\
 & Medium, second & $ 0.81 \pm  0.00$ & $ 0.239 \pm  0.003$ & $ 4.15 \pm  0.03$ & $ 0.750 \pm  0.012$ \\
 & Brightest, total & $20.64 \pm  0.10$ & $ 0.448 \pm  0.006$ & $143.55 \pm  1.14$ & $ 1.077 \pm  0.017$ \\
 & Brightest, second & $ 3.90 \pm  0.03$ & $ 0.957 \pm  0.013$ & $37.46 \pm  0.33$ & $ 1.465 \pm  0.021$ \\
\hline
Moderately prolate & Unbiased & $ 0.56 \pm  0.00$ & $ 0.010 \pm  0.000$ & $ 1.77 \pm  0.00$ & $ 0.035 \pm  0.001$ \\
 & Faintest, total & $ 1.14 \pm  0.00$ & $ 0.020 \pm  0.000$ & $ 4.49 \pm  0.01$ & $ 0.057 \pm  0.001$ \\
 & Faintest, second & $ 0.74 \pm  0.00$ & $ 0.021 \pm  0.000$ & $ 3.07 \pm  0.01$ & $ 0.058 \pm  0.001$ \\
 & Medium, total & $ 2.06 \pm  0.00$ & $ 0.025 \pm  0.001$ & $ 8.92 \pm  0.02$ & $ 0.065 \pm  0.001$ \\
 & Medium, second & $ 1.01 \pm  0.00$ & $ 0.032 \pm  0.001$ & $ 3.07 \pm  0.01$ & $ 0.058 \pm  0.001$ \\
 & Brightest, total & $25.61 \pm  0.09$ & $ 0.064 \pm  0.001$ & $172.87 \pm  0.56$ & $ 0.109 \pm  0.002$ \\
 & Brightest, second & $ 4.78 \pm  0.03$ & $ 0.166 \pm  0.003$ & $45.60 \pm  0.23$ & $ 0.203 \pm  0.003$ \\
\hline
Very prolate & Unbiased & $ 0.57 \pm  0.00$ & $ 0.022 \pm  0.000$ & $ 1.83 \pm  0.01$ & $ 0.078 \pm  0.001$ \\
 & Faintest, total & $ 1.15 \pm  0.00$ & $ 0.040 \pm  0.001$ & $ 4.60 \pm  0.01$ & $ 0.117 \pm  0.002$ \\
 & Faintest, second & $ 0.74 \pm  0.00$ & $ 0.042 \pm  0.001$ & $ 3.13 \pm  0.01$ & $ 0.117 \pm  0.002$ \\
 & Medium, total & $ 2.07 \pm  0.01$ & $ 0.049 \pm  0.001$ & $ 9.11 \pm  0.03$ & $ 0.131 \pm  0.002$ \\
 & Medium, second & $ 1.00 \pm  0.00$ & $ 0.062 \pm  0.001$ & $ 5.19 \pm  0.02$ & $ 0.142 \pm  0.002$ \\
 & Brightest, total & $25.51 \pm  0.10$ & $ 0.119 \pm  0.002$ & $174.92 \pm  0.64$ & $ 0.210 \pm  0.003$ \\
 & Brightest, second & $ 4.59 \pm  0.03$ & $ 0.304 \pm  0.006$ & $45.29 \pm  0.25$ & $ 0.370 \pm  0.005$ \\
\hline
Triaxial & Unbiased & $ 0.51 \pm  0.00$ & $ 0.042 \pm  0.000$ & $ 1.55 \pm  0.00$ & $ 0.176 \pm  0.002$ \\
 & Faintest, total & $ 1.03 \pm  0.00$ & $ 0.073 \pm  0.001$ & $ 3.86 \pm  0.01$ & $ 0.260 \pm  0.002$ \\
 & Faintest, second & $ 0.66 \pm  0.00$ & $ 0.077 \pm  0.001$ & $ 2.60 \pm  0.01$ & $ 0.261 \pm  0.002$ \\
 & Medium, total & $ 1.86 \pm  0.00$ & $ 0.089 \pm  0.001$ & $ 7.62 \pm  0.02$ & $ 0.292 \pm  0.002$ \\
 & Medium, second & $ 0.90 \pm  0.00$ & $ 0.113 \pm  0.001$ & $ 4.27 \pm  0.01$ & $ 0.321 \pm  0.003$ \\
 & Brightest, total & $23.07 \pm  0.06$ & $ 0.218 \pm  0.002$ & $144.19 \pm  0.44$ & $ 0.490 \pm  0.004$ \\
 & Brightest, second & $ 3.96 \pm  0.02$ & $ 0.636 \pm  0.006$ & $35.99 \pm  0.14$ & $ 0.983 \pm  0.007$ \\
\hline
Mixed & Unbiased & $ 0.49 \pm  0.00$ & $ 0.059 \pm  0.001$ & $ 1.47 \pm  0.00$ & $ 0.220 \pm  0.002$ \\
 & Faintest, total & $ 0.98 \pm  0.00$ & $ 0.099 \pm  0.001$ & $ 3.69 \pm  0.01$ & $ 0.316 \pm  0.003$ \\
 & Faintest, second & $ 0.63 \pm  0.00$ & $ 0.103 \pm  0.001$ & $ 2.49 \pm  0.01$ & $ 0.317 \pm  0.003$ \\
 & Medium, total & $ 1.77 \pm  0.00$ & $ 0.118 \pm  0.001$ & $ 7.28 \pm  0.02$ & $ 0.354 \pm  0.003$ \\
 & Medium, second & $ 0.86 \pm  0.00$ & $ 0.148 \pm  0.001$ & $ 4.10 \pm  0.01$ & $ 0.385 \pm  0.003$ \\
 & Brightest, total & $22.24 \pm  0.06$ & $ 0.281 \pm  0.003$ & $139.49 \pm  0.46$ & $ 0.582 \pm  0.005$ \\
 & Brightest, second & $ 3.94 \pm  0.02$ & $ 0.752 \pm  0.007$ & $35.64 \pm  0.14$ & $ 1.103 \pm  0.008$ \\
\hline
\end{tabular}
\end{center}
\end{table*}

\subsubsection{Total cross-section}

We first consider how the orientation-averaged unbiased total
cross-section varies with model shape for the lower mass model in
Table~\ref{T:haloshape}.  When progressing from spherical to
moderately and very oblate shapes, $\avst$ decreases by 8 and 12 per
cent, respectively.  When progressing from spherical to moderately and
very prolate shapes, $\avst$ increases by 6 and 8 per cent,
respectively.  The triaxial model has $\avst$ that is 3 per cent lower
than in the spherical case.  Finally, the cross-section for the mixed
shape model is the same as that for the moderately oblate model, which
suggests that the oblate stellar component is dominant in determining
the lensing properties, even in the mixed model.
Table~\ref{T:haloshape} shows that these trends are only slightly
enhanced for the higher mass model.  We conclude that the
orientation-averaged unbiased cross-section does \emph{not}
depend strongly on the intrinsic model shape, even for significant
changes in shape.  In other words, while there is an orientation
bias such that non-spherical lens galaxies are more likely to be
aligned along the line-of-sight, there is no shape bias that
would make non-spherical galaxies significantly over- or
under-abundant in strong lens samples.

Once we include magnification bias, the trends with model shape tend
to be somewhat stronger, particularly when the survey flux limit is in
the bright (steep) part of the luminosity function.  However, even in the
worst case, the variations in $\avst$ with model shape are
a maximum of $\sim 30$ per cent, rather than $\sim 10$ per cent in the
unbiased case.  This is clearly not a selection bias that will
drastically alter the distribution of model shapes for strong lensing
systems from the distribution of model shapes for all galaxies at that
mass.  Furthermore, while we have only tested specific sets of
intrinsic axis ratios for the non-spherical shapes, the similarity of
the cross-sections to the spherical case suggests that at least for
axis ratios $b/a$ and $c/a$ equal to or greater than the cases we have
tested, our conclusions still apply.

With Table~\ref{T:haloshape} we can consider how the shape of the
galaxy affects the mass bias (at least for $\gdm=1$; recall that
Section~\ref{SS:basicresults} shows how the mass bias depends on $\gdm$ for
spherical models). For all model shapes, the unbiased total
cross-section increases by a factor of $\sim 3$ from the lower to the
higher mass scale.  The difference is even larger when considering
biased cross-sections, up to a factor of $\sim 9$ for cases with a
flux limit in the steepest part of the luminosity function.  We infer
that mass bias is not very sensitive to galaxy shape (which is not
surprising given that there is little dependence of
$\avst$ on shape).  However, mass bias is sensitive to
magnification bias such that in surveys with flux limits on the
steeper part of the quasar luminosity function, the observed strong
lensing systems will be preferentially skewed towards higher masses
even more so than for those surveys with fainter flux limits.

Our results compare well with those in the literature.
\citet{1998ApJ...495..157K} noted that there is a significant
orientation bias for strong lensing by spiral galaxies, but found that
the orientation-averaged cross-sections of disk+halo models differ
from those of spherical models by $\lesssim 10$ per cent.
\citet{2005ApJ...624...34H}\footnote{ \citet{1996ApJ...473..595K},
  \citet{1997ApJ...482..604K}, \citet{2001ApJ...553..709R}, and
  \citet{2003MNRAS.346..746C} also computed lensing cross-sections for
  isothermal ellipsoid lens models with different ellipticities, but
  those studies did not isolate and discuss the effects of shape on
  the total cross-section in the same way that
  \citet{2005ApJ...624...34H} did.} used single-component projected
isothermal ellipsoid models and found that, when the Einstein radius
is fixed, the ellipticity affects lensing cross-sections at the few
per cent level.  One instance of a shape bias was found by
\citet{2004ApJ...610..663O}, who found that $\avst$
increases with triaxiality for massive, generalised NFW halos.
However, they noted that the increase was driven by the cross-section
for three-image lenses---an effect that was enhanced by the use of a
single-component generalised NFW dark matter halo without a stellar
component, such that the inner density profile was relatively shallow
and hence quite prone to create three-image lens systems.
Such models may or may not be reasonable for the massive halos of
galaxy clusters, but are certainly not realistic models for galaxies.
Even so, it is interesting to note that \citet{2004ApJ...610..663O}
found the net cross-section for two- and four-image lenses to be
roughly independent of halo shape. Altogether, we conclude that
there is general consensus that shape bias is not a strong effect in
strong lensing by galaxies.

Finally, we note that our conclusions about shape bias are not a
strong function of mass model parameters such as $\gdm$.  This
conclusion is expected given our agreement with other papers that
use quite different mass models. However, as in Section~\ref{SS:basicresults}
and here, our conclusion regarding the degree of mass-based selection
bias does depend on $\gdm$ and the magnification bias mode.

\subsubsection{Quad/double ratio}\label{SS:quaddouble}

Next, we consider the ratio of the orientation-averaged four-image
and two-image cross-sections to understand the relative abundances
of quad and double lenses.  Table~\ref{T:haloshape} lists
$\avsig{4}/\avsig{2}$ for all shape models, for both mass scales.
Naturally, in the spherical case this ratio is identically zero.  For
oblate models, the abundance of four-image lens systems can be quite
significant, particularly when the flux limit is in the steep part of
the luminosity function: for example, for the very oblate, higher mass
model the quad/double ratio ranges from $\sim 0.5$ in the unbiased
case to $\sim 1.5$ for some of the magnification bias modes.  Prolate
systems, in contrast, have fairly low quad/double ratios, reaching a
maximum of $37$ per cent. The trend with model shape is sensible: for
the moderately and very oblate cases the smallest axis ratio $\cadm$
is (respectively) $0.50$ and $0.36$, whereas in the moderately and very
prolate case the smallest axis ratio $\badm = \cadm$ is (respectively)
$0.71$ and $0.60$.  Thus, the maximum projected ellipticity is lower
in the prolate case, so there are fewer four-image lenses.

The trend with magnification bias is due to the fact that
four-image systems have preferentially higher magnifications
than two-image systems.  Moreover, four-image lenses often have
two or more particularly bright images, which is why the
quad/double ratio is especially high if we compute magnification
bias using the second-brightest image (see also
Fig.~\ref{F:incoblpro}).  In the triaxial case, the quad/double
ratio is particularly dependent on how magnification bias is
treated, ranging from $0.04$ to $1.00$.

Generically, for a given shape model we find that $\avsig{4}/\avsig{2}$
is larger for our higher mass model, because at the higher mass scale
the dark matter component (which is flatter than the stellar component)
plays a more important role.  Our conclusion is seemingly at odds with
that of \cite{2007MNRAS.379.1195M}, who found that the four-image systems
in their simulations are preferentially at lower velocity dispersion
than the two-image systems.  However, \cite{2007MNRAS.379.1195M} also
found that the four-image systems have lens galaxies with a more
pronounced disk component, in which case the appropriate comparison is
not at fixed DM halo shape. Thus, there is no real disparity between
our results and theirs, since we find that oblateness (as would occur
with a pronounced disk component) strongly increases
$\avsig{4}/\avsig{2}$.

Our results are compatible with the conventional wisdom that the
quad/double ratio in strong lensing offers a valuable probe of the
density shapes in the inner regions of galaxies
\citep[e.g.,][]{2001ApJ...553..709R,2007NJPh....9..442O}.  But
magnification bias also plays a very important role \citep[see
also][]{2005ApJ...624...34H,2007MNRAS.379.1195M}.  In particular,
\citet{2007NJPh....9..442O} suggests that magnification bias may
explain why there is a higher fraction of four-image lenses in radio
surveys than in optical surveys.  He suggests that the source
luminosity function is steeper in radio surveys than in optical
surveys (at least, in optical surveys that probe deeper than the break
in the quasar luminosity function), which enhances magnification bias
and (as we have seen) increases the quad/double ratio in radio
surveys.

\subsubsection{Three-image cross-section}

The final shape-related question we address is which shapes give rise
to three-image lens systems.  They can only occur in cases of such
high projected ellipticity that the tangential caustic protrudes beyond
the radial caustic. For the lower-mass model, we find three-image
systems only for the very oblate model, with a cross-section that
depends sensitively on the dark matter inner slope. This dependence
may arise from the fact that our dark matter is flatter than the
stellar components, and models with steeper dark matter inner slope
are more dark matter-dominated in the region where strong-lensing is
important.  Thus, for example, the three-image cross-section
(unbiased) for this very oblate lower mass model averages to zero for
$\gdm = 0.5$ and $\gdm = 1$, and is small but non-zero for $\gdm =
1.5$ (but $\avsig{3}$ is still just 0.3 per cent of \avst).
Magnification bias can enhance the relative abundance of three-image
systems by up to a factor of $\sim 10$ depending on the survey flux
limit (with the greatest enhancement when the flux limit is on the
bright end). Naturally these systems can only occur in the oblate case
when the galaxy is seen close to edge-on, which is a fairly extreme
orientation-related selection bias inherent to such systems.

For the higher-mass model, three-image lens systems occur for both the
moderately and very oblate cases (presumably because the higher-mass
model is inherently more dark matter-dominated and therefore flatter);
and there is again a strong trend with $\gdm$.  For the very oblate
shape, the unbiased three-image lens system abundance
($\avsig{3}/\avst$) ranges from 0.7 to 8 per cent as \gdm\ changes
from 0.5 to 1.5, a factor of 10 increase.  For the most extreme
magnification bias mode ($4L_*$ with the total magnification), the
increase is from $\sim 3$ per cent at $\gdm=0.5$ to $\sim 20$ per cent
at $\gdm=1.5$ (a factor of 7).  An observational scarcity of
three-image lens systems with group-mass lenses can therefore
constrain the abundance of strongly oblate matter distributions (but
see below).  For this mass scale, the triaxial and mixed models also
give rise to three-image lenses, but with a relative abundance that is
below 1 per cent for all $\gdm$ and magnification bias modes.

We contrast this result with the findings in
\citet{2004ApJ...610..663O}, who use single-component generalised NFW
profiles with an axis ratio distribution (rather than fixed shapes as
used here) to study the trends of $\avsig{3}/\avst$ with \gdm.  That
work suggests that the three-image abundance increases as $\gdm$ gets
\emph{shallower}, which seems to contradict our trend.  The difference
between these two results stems from a difference in modeling
assumptions: in their case, without a fixed stellar component creating
a lower bound to the central density, decreasing \gdm\ strongly
affects the (lens-plane) inner critical curve (which is very sensitive
to the central slope of the density profile), shrinking the
(source-plane) radial caustic.  However, the tangential caustic does
not shrink as much, because it depends primarily on the projected
ellipticity.  As a result, decreasing \gdm\ actually increases the
area that is inside the tangential caustic and outside the radial one,
where three-image systems originate.  In contrast, when we decrease
\gdm\ the radial caustic does not shrink very much, since we have a
fixed stellar component.  However, the stellar component becomes much
more dominant, and since we have modeled it as being rounder, this
means the tangential caustic will shrink.  The result is that the area
inside the tangential caustic and outside the radial one, and
therefore the three-image lensing system abundance, is reduced.

This comparison suggests that the predicted abundance of three-image
strong lenses is quite dependent on modeling assumptions,
particularly on the ratio of stellar to dark matter mass and 
the assumed shapes for the two components.  Another potential
concern in studying the statistics of three-image lenses is that
a four-image lens could be misclassified as a three-image system
if one of the images is too faint to be detected.  Since typically
$\avsig{4} \gg \avsig{3}$, any small rate of contamination from
misclassified four-image lenses could corrupt the three-image
sample.  In practise, the frequency with which such errors may
occur will depend sensitively on the galaxy model and on the
survey parameters, so a detailed exploration is beyond the scope
of this paper.  For now, we conclude that the abundance of
three-image lenses contains interesting information about the
abundance of very oblate galaxies, but there may be practical
challenges in extracting that information from observed lens
samples.

%


\subsection{Magnification bias}
\label{SS:resultsmagbias}

As an aside, we now investigate whether the conversion from unbiased
to the various biased cross-sections can be quantified with a single
number for each magnification bias mode, independent of mass, shape,
dark matter inner slope, or the number of images.  This would allow
considerable simplification, not only in the presentation of our
results, but also in future calculations.  We again refer to the
orientation-averaged cross-sections in Table~\ref{T:haloshape}.

First, we find that for $\gdm = 1$ (as in the table), and for a given
mass scale, the conversion from unbiased to biased $\st$ is relatively
independent of shape.  This statement is true within $\sim 2$ per cent
for the fainter flux limits, or within $\lesssim 10$ per cent for the
brightest flux limit. However, the conversion does depend strongly on
$\gdm$ and mass. In retrospect, this is relatively easy to understand,
because the biased cross-section depends not just on the lensing
caustics, but also on the lensing magnification as a function of
position, which in turn depends on how the projected density is
distributed within the Einstein radius.  Changing the dark matter
inner slope changes both the caustics and the magnification
distribution, and not necessarily in the same way.  Likewise, the
different mass scales have different balances of stellar and dark
matter components, each with its own density profile, so the
conversion from unbiased to biased cross-sections is not the same for
our lower and higher mass scales.

Furthermore, because the four-image cross-section depends on the
projected shape (Section~\ref{SS:resultsshape}), the conversion of the
quad/double ratio from the unbiased to the biased cases depends not
just on $\gdm$ and the mass, but also on the shape. We therefore
conclude that for the mass models used here, there is no simple
conversion from unbiased to biased cross-sections that would allow us
to simplify the presentation of our results. Selection biases in $\st$
and $\sig{4}/\sig{2}$ with the form of the mass density profile may
depend strongly on the survey detection strategy and flux limit.
Consequently, we continue to present results for the different
magnification bias modes throughout the rest of this paper.

\section{Inner slope}
\label{S:innerslope}

We now examine the role of the galaxy density profile in more detail.
We study how the dark matter inner slope \gdm\ affects
orientation-averaged cross-sections (Section~\ref{SS:resultsgamma}) as well
as lens image separation distributions
(Section~\ref{SS:resultsimagesep}).  In Section~\ref{SS:resultsAC} we extend the
discussion of profile-dependent selection biases to consider the
effect of adiabatic contraction on the strong lensing cross-section
(for spherical models only).

\subsection{Inner slope}
\label{SS:resultsgamma}

\begin{figure}
\begin{center}
  \includegraphics[width=\columnwidth,angle=0]{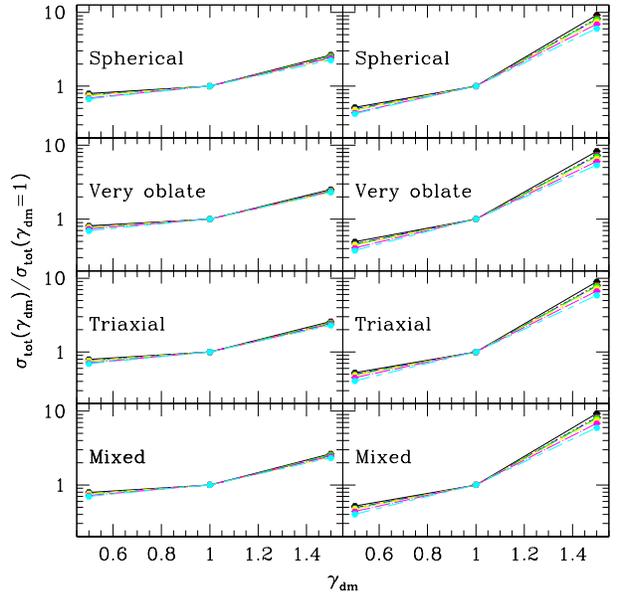}
  \caption{\label{F:gammacs} Variation of total, orientation-averaged
    lensing cross-section with $\gdm$ for the lower (left) and higher
    (right) mass scales for different shape models as labeled on the
    plot, with the magnification bias mode indicated as in
    Fig.~\ref{F:incoblpro}.}
\end{center}
\end{figure}

\begin{figure}
\begin{center}
\includegraphics[width=\columnwidth,angle=0]{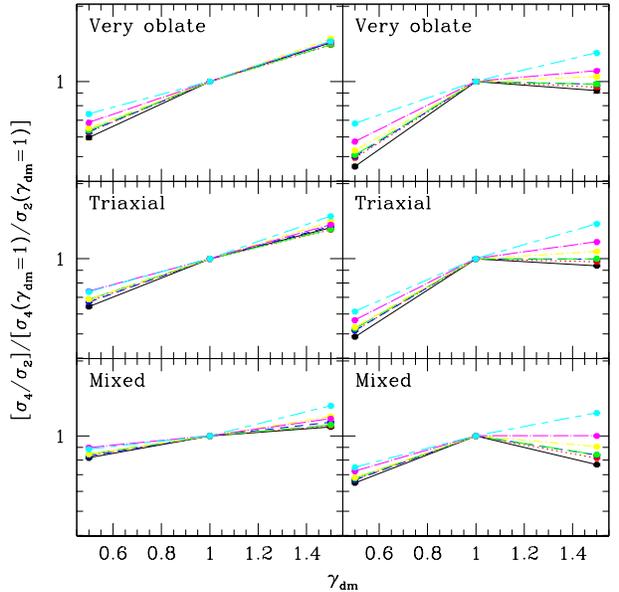}
\caption{\label{F:gamma42} Variation of the orientation-averaged
  quad/double ratio with $\gdm$ for the lower (left) and higher
  (right) mass scales for different shape models as labeled on the
  plot, with magnification bias mode indicated as in
  Fig.~\ref{F:incoblpro}.}
\end{center}
\end{figure}

In the sections that follow, we typically use orientation-averaged
cross-sections.  While in sections that dealt with orientation and halo
shape, we have introduced the notation
$\avst$ to highlight the difference between cross-sections for a
single orientation versus 
orientation-averaged  cross-sections, we now revert to simply using 
$\st$ for orientation-averaged cross-sections since they are now our sole focus.

Fig.~\ref{F:gammacs} shows the variation of the total cross-section
with the dark matter inner slope $\gdm$ for both mass scales, for four
of the seven shapes and all seven magnification bias modes.  In order
to highlight the variation with $\gdm$ (and to fit all results for a
given shape in a single panel), we normalise $\st$ for each
magnification bias mode by the corresponding values when $\gdm=1$
(given in Table~\ref{T:haloshape}). Statistical error bars
are shown on the plot, but they are generally smaller than the sizes
of the points themselves due to the large number of Monte Carlo
realisations.  Fig.~\ref{F:gamma42} shows analogous results for the
quad/double ratios (once again, normalised by the values at $\gdm=1$
given in Table~\ref{T:haloshape}).  We again show different shape
models, although we eliminate the spherical model since its four-image
cross-section is identically zero.

The total lensing cross-section for the lower mass model
(Fig.~\ref{F:gammacs}, left column) shows several interesting trends
with $\gdm$. First, as we expect, $\st$ increases with $\gdm$ because
the projected mass density is higher in the inner parts for steeper
$\gdm$. Second, all the curves for a given mass scale are nearly the
same, with $\st$ decreasing by 20 to 30 per cent when going from
$\gdm=1$ to $\gdm=0.5$, and increasing by a factor of $\sim 3$ when
going from $\gdm=1$ to $\gdm=1.5$.  The variation within the quoted
ranges depends only slightly on the magnification bias mode, and is
nearly independent of the shapes considered here.  For the higher mass
scale (Fig.~\ref{F:gammacs}, right column), we likewise see that the
results are consistent for different shapes.  At this mass scale, \st\
decreases by 50--70 per cent going from $\gdm=1$ to $\gdm=0.5$, and it
increases by factors of 6--10 going from $\gdm=1$ to $\gdm=1.5$ (where
the quoted range encompasses the different magnification bias modes).
The reason the dark matter inner slope causes more significant selection bias for the
higher mass model is that the flatter dark matter is more prominent in
that model.

To understand the strong variation with $\gdm$ further, we explore a
finer grid in $\gdm$ in Section~\ref{SS:resultsconcgamma} below, and we
compare the variation in the cross-section with the variation in the
projected dark matter fraction.  However, it is apparent from 
paper~I figure 6 
that the projected mass density
(and therefore the lensing properties) is dominated on small scales by
the stellar component for $\gdm=0.5$ and $\gdm=1$, whereas the dark
matter component has a significant contribution everywhere for
$\gdm=1.5$.  Consequently, the cross-section variation is quite likely
related to the change in the projected dark matter fraction.

Next, we consider the quad/double ratio for the lower mass model in
the left column of Fig.~\ref{F:gamma42}. For all shapes shown here,
this ratio is roughly exponential in $\gdm$, $\sig{4}/\sig{2} \propto
\exp{(0.85\gdm)}$, once again roughly independent of model shape and
with consistent trends for all magnification bias modes. Moving to the
higher mass scale in the right column, we see rather different
results: the quad/double ratio increases from $\gdm=0.5$ to $\gdm=1$,
but then approximately flattens out from $\gdm=1$ to $\gdm=1.5$.
There is actually a slight decrease in $\sig{4}/\sig{2}$ for the
unbiased cross-section, and a slight increase for the significant
magnification bias modes. This result involves a combination of
several effects, the sizes of which approximately cancel, and which
are complicated by the fact that we are working with
orientation-averaged quantities. First, examination of the mock lens
systems suggests that the radial caustic moves strongly outwards for
$\gdm=1.5$ for the higher mass scale (leading to the very large
increase in \st\ shown in Fig.~\ref{F:gammacs}).  The tangential
caustic increases in size somewhat, because it depends predominantly
on the projected shape (which is flatter at high $\gdm$ due to the
dominant DM component), so while $\sig{4}$ increases, so does $\sig{2}
\approx \st-\sig{4}$, thus leading to a flattening out of
$\sig{4}/\sig{2}$.  Because the increase in \st\ is not nearly as much
for the lower mass model when going from $\gdm=1$ to $1.5$ (due to the
lower dark matter fraction), $\sig{4}/\sig{2}$ does not level out as
for the higher mass model.  The trend with magnification bias is also
understandable: for more significant magnification bias (higher flux
limit and steeper luminosity function), the inner regions where
$4$-image systems dominate become more important, which is why for the
higher mass model the biased $\sig{4}/\sig{2}$ can still increase
slightly when going from $\gdm=1$ to $1.5$ even though the unbiased
quad ratio decreases.

Several previous studies \citep[e.g.,][]{2000ApJ...532..679P,
  2001ApJ...549L..25K, 2001ApJ...555..504W, 2002ApJ...566..652L,
  2004ApJ...600L...7H} used spherical single-component mass models to
examine how the lensing cross-section depends on the inner slope of
the density profile.  \citet{2001ApJ...549L..25K} found in particular
that the cross-section variation could be explained mainly by changes
in the fraction of the total mass that lies within some fiducial
radius.  The simple scaling with the ``core mass fraction'' is
probably broken in our models by the presence of the stellar
component, although we do examine whether there is some analogous
scaling in Section~\ref{SS:resultsconcgamma} below.  There is general
consensus that the total cross-section can increase fairly
dramatically as the inner density profile steepens, although we
believe our two-component models better isolate the effects of the
dark matter slope alone.

In conclusion, there is a strong selection bias in the total
cross-section with \gdm\ that can amount to factors of a few
(particularly within the upper end of the allowed range); and this
bias is remarkably insensitive to the model shape and only mildly
sensitive to the magnification bias (survey flux limit and detection
strategy).  The selection bias with \gdm\ in the quad/double ratio
can amount to several tens of per cent, and is likewise relatively
insensitive to the model shape and magnification bias mode. 

In this section, we have focused on quantities averaged over
orientation.  In Appendix~\ref{A:gammaorient}, we indicate how the
variation of cross-section with $\gdm$ depends only mildly on
orientation.  This result is relevant only when a special subset
of strong lenses are selected based on their projected shape.

\subsection{Image separation distributions}\label{SS:resultsimagesep}

We next consider the effects of the dark matter inner slope $\gdm$
on the image separation distribution, and consider whether this is
a magnification bias-dependent quantity.  This calculation is done
separately in the spherical and the non-spherical cases.

\subsubsection{Spherical models}

We begin with cusped, spherical density profile models, which lead
only to strong lensing systems with two images.  
We first consider what to expect for the image separation distribution
based on general considerations.  By definition, the Einstein radius
\Rein\ for a circularly-symmetric projected mass distribution is the
radius at which the lensed image is a ring because the source is
located precisely behind the lens.  In that case, the image separation
is $\imsp=2\Rein$.  Consequently, we expect that regardless of
the form of the density profile, it must approach $2\Rein$,
because the sources near the source plane origin produce precisely that image
separation.  

The simplest comparison
we can make is against the very simple image separation
distribution for the SIS density profile, which is a delta function at $2\Rein$:
\begin{equation}
p_{\mathrm{SIS}}(\imsp)=\delta(\imsp-2\Rein).  
\end{equation}
We ask, then, how the complexity of our mass models causes the
image separation distribution to deviate from this simple form.  

\begin{figure}
\begin{center}
  \includegraphics[width=\columnwidth,angle=0]{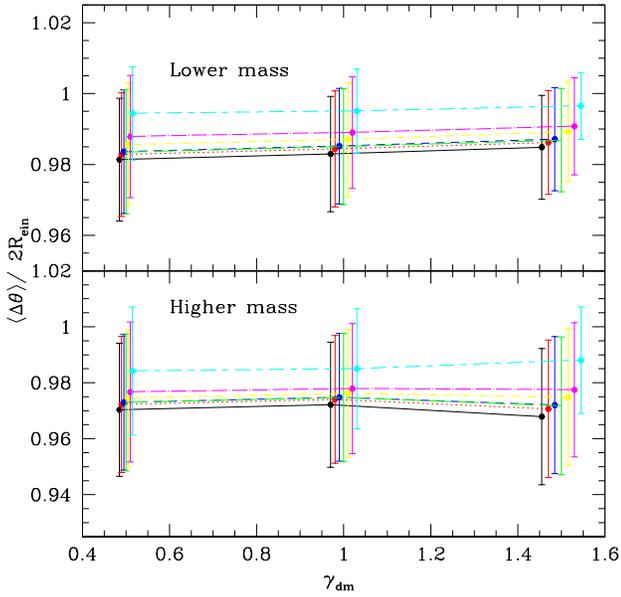}
  \caption{\label{F:sphimgsep}Mean (points) and width (errorbars) of
    the image separation distribution for the spherical shape model,
    in units of $2\Rein$, as a function of $\gdm$.  Points are
    horizontally offset for clarity, but all measurements are with the
    values of $\gdm=\{0.5,1,1.5\}$.  Line colours and types indicate
    the magnification bias mode, as described in the caption for
    Fig.~\ref{F:incoblpro}.}
\end{center}
\end{figure}

Fig.~\ref{F:sphimgsep} shows the mean image separation in units of
twice 
the Einstein radius as a function of $\gdm$, for all magnification
bias modes.  The standard deviation of the image separation
distribution is shown as an errorbar.  
There are a few conclusions we can draw from this plot.  First, the
image separation distribution depends on $\gdm$ almost exclusively
through its dependence on $\Rein$.  Second, despite the complexity of
the mass models, to lowest order the distributions are remarkably
close to the distributions for a pure SIS, very narrow and with an
average that is only $\sim 2$--$3$ per cent below $2\Rein$. Finally,
there is a minor (for practical purposes) but nonzero dependence on
the magnification bias mode.  The mean of the image separation
distribution increases with more significant magnification bias,
approaching $2\Rein$ from below, because the magnification is highest
for a source near the origin, where $\imsp$ is precisely $2$\Rein.

We have quantified the image separation statistics using the
usual mean and standard deviation.  However, the distributions
$p(\imsp)$ are not remotely Gaussian or even peaked at $\avimsp$;
rather, they are steeply increasing functions of $\imsp$ that
are truncated abruptly at $\imsp=2\Rein$.  We use 
the deflection angle profile $\alpha(R')$ in paper~I figure 7 
to explain the shape of this
distribution, and why the mean is so close to the SIS result.  As
shown there, for the three cusped models, $\alpha(\theta)$ is nearly
flat and equal to $\Rein$ from $R' \sim \Rein/4$ until well above
$\Rein$.  Consequently, the image separations should be below
$2$\Rein\ for those small subset of images that appear to be near the
origin in the lens plane, but approach $2\Rein$ for the majority of
the images.

Our results are consistent with those of \cite{2002ApJ...568..488O},
who study image separations for spherical, single-component,
generalised NFW models, and also find that it is close to, but
slightly below, $2\Rein$.  In light of that result, ours is not
surprising, since the superposition of a stellar component with a
generalised NFW model tends to yield a composite model that is closer
to isothermal near $\Rein$ and therefore more likely to give
quasi-isothermal image separation distributions.

Thus, for spherical models, the changes in the image separation
distribution due to changes in density profile parameters (mass, \gdm)
arise almost exclusively through the change in the Einstein radius.
This is a remarkably simple result given the complexity of the mass
models used.  We note, however, that while the physical scaling of the
image separation distribution is simple, there may be observational
selection biases due to resolution limits (more likely to miss lenses
with smaller \Rein) or due to finite apertures for the image search
(more likely to miss lenses with larger \Rein).  To estimate the
effects of selection bias in any given survey, it will be necessary to
supplement the physical selection effects we study here with an
analysis of observational selection effects.

\subsubsection{Non-spherical models}

\begin{figure}
\begin{center}
  \includegraphics[width=\columnwidth,angle=0]{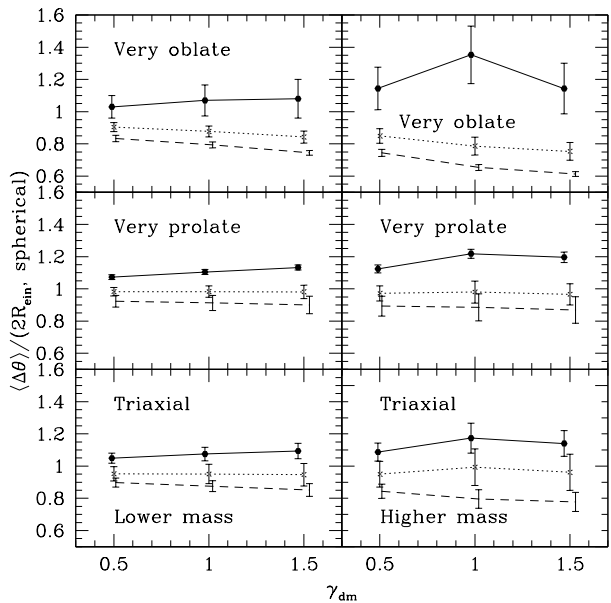}
  \caption{\label{F:nonsphimgsep} Mean (points) and width (errorbars)
    of the unbiased image separation distribution, relative to
    $2$\Rein\ for the spherical shape model, as a function of $\gdm$.
    The left and right panels show the lower and higher mass scales,
    respectively.  Points are horizontally offset for clarity, but all
    measurements are with the values of $\gdm=\{0.5,1,1.5\}$. The top
    panels show the very oblate model shape, where the solid line =
    edge-on, dotted line = intermediate, and dashed line = face-on
    configuration.  The middle panels show the very prolate model
    shape, where the solid line = end-on, dotted line = intermediate,
    and dashed line = side-on configuration.  The bottom panels show
    the triaxial model shape, where the solid line = viewed with line-of-sight along $a$,
    dotted line = viewed along $b$, and dashed line = viewed along
    $c$.}
\end{center}
\end{figure}

For non-spherical models, we have defined the image separation as the
maximum separation between any pair of images in a given image system,
regardless of the total number of images.
We have then calculated the mean and width of the image separation
distribution as a function of $\gdm$ for both mass scales, with
different extreme configurations in the projected shapes.  These
results are shown for the unbiased case in 
Fig.~\ref{F:nonsphimgsep}, normalised by twice the spherical
Einstein radius (i.e., we use the same normalisation factors that
went into Fig.~\ref{F:sphimgsep}).  As for the spherical case, the
effects of magnification bias (not shown) are subtle. 

There are a few conclusions we can draw from this plot.  First, as
expected, the mean of the image separation distribution scales with
$\gdm$ in nearly the same way as the Einstein radius in the spherical
case, so the curves are nearly flat.  Second, the difference between
the means and widths of $p(\imsp)$ for various shape configurations
can be understood simply in terms of the projected shape.  For
example, for the very oblate and very prolate shapes, the solid lines
(edge-on and end-on, respectively) give the highest \emph{mean}
$\imsp$ because they give a larger projected mass density for a given
(azimuthally-averaged) separation from the centre of mass than in
configurations represented with dashed lines (face-on and side-on,
respectively), consistent with the larger total lensing cross-section.
Because the projected ellipticity is higher for the solid line
configuration in the oblate case (edge-on) and for the dashed line in
the prolate case (side-on), the \emph{width} of the image separation
distributions in those cases is larger (depending on where the source
is located in the region where lensing occurs, very different image
configurations can occur).  In the triaxial case, these arguments
imply that the mean image separation is highest when viewing along the
major axis (solid lines), but the width is highest when viewing along
the intermediate axis (dotted lines).  When considering the image
separation distribution derived from many lenses drawn from galaxies
of a given shape with random orientations, we must average the image
separation distribution for each orientation weighted by the
cross-section \st\ for that orientation. We have confirmed that the
differences between $\avimsp$ for different orientations relative to the 
spherical Einstein radius can be accounted for by computing $\Rein$
from the monopole deflection (equation~26 of Paper~I) as $\alpha_0(\Rein) =
\Rein$.

Our finding that non-spherical models have significantly broader image
separation distributions ($\sim 10$ per cent rather than a few per
cent), but mean values that are not strongly different from the
spherical case, is consistent with the results in
\cite{2005ApJ...624...34H}, who tested this effect using
single-component singular isothermal ellipsoid models.  The fact that
the image separation distribution is quite close to that for a
spherical isothermal model even for these highly non-spherical and
non-isothermal models is quite remarkable.  The implication is that
there may be an observational selection bias that is sensitive to
\gdm, depending on the angular resolution and aperture limits for a
particular survey.  For example, an inability to find large-separation
lens systems would tend to counteract the physical selection bias
towards galaxies with steeper dark matter inner slope ($\gdm$),
particularly at low redshift.

\subsection{Adiabatic contraction}
\label{SS:resultsAC}

As discussed in Paper~I, the need for adiabatic contraction of dark
matter halos in response to the adiabatic infall of baryons to form a
galaxy \citep{1980ApJ...242.1232Y, 1986ApJ...301...27B,
  2004ApJ...616...16G, 2005ApJ...634...70S} is unclear for early-type
galaxies.  In this subsection, we test the effects of adiabatic
contraction explicitly, for spherical, cusped $\gdm=1$ models only.
\citet{2001ApJ...559..531K} previously examined the effect of AC on
the strong lensing image separation distribution, and
\citet{2001ApJ...561...46K} included AC when computing lens
statistics, but neither study considered it as a selection bias as we
are.

The effects of AC on the projected density profiles are shown in 
paper~I figures 4--7; it increases the dark
matter density on scales below $\sim 10$ kpc. 
Due to the increased density in the inner
regions, we expect an increase in the strong-lensing cross-section.
Our purpose is to determine the significance of this increase compared
to the increase when changing $\gdm=1$ to $\gdm=1.5$ without AC.

For the lower mass model, the total lensing cross-section $\st$
increases by a factor of $2.6$ when going from $\gdm=1$ to $\gdm=1.5$,
or slightly less than that for the higher magnification bias modes. In
contrast, when going from $\gdm=1$ without AC to including AC, the
cross-section increases by a factor of $\sim 2.6$ for the
\cite{1986ApJ...301...27B} prescription, or a factor $\sim 2.0$ for
the \cite{2004ApJ...616...16G} prescription.  These factors are
reduced to $\sim 2.0$ and $\sim 1.5$ for the higher magnification bias
modes.

For the higher mass model, the unbiased $\st$ increases by a factor of
$9$ when going from $\gdm=1$ to $\gdm=1.5$, or as little as $6$ for
the higher magnification bias modes. In contrast, when going from
$\gdm=1$ without AC to including AC, $\st$ increases by a factor $\sim
3.5$ for the original \cite{1986ApJ...301...27B} prescription, or a
factor $\sim 2.3$ for the \cite{2004ApJ...616...16G} prescription.
These factors are again reduced for the higher magnification bias
modes.  For comparison, previous studies of the effect of baryonic cooling on strong
lensing by \emph{clusters} also found that the cross-section increases
by a factor of a few \citep[e.g.,][]{2005A&A...442..405P,
  2008ApJ...687...22R,2008ApJ...676..753W}.

We conclude that AC has typically a less significant effect on
the strong lensing cross-section than increasing $\gdm$ from $1$ to
$1.5$ for both mass scales. Nonetheless, the increases in the strong
lensing cross-section due to AC are non-negligible.  There are several
implications of this finding. First, if AC happens in typically all
galaxies of these types (i.e., the assumptions behind it are valid and
the changes in profile are preserved through mergers), it suggests
that the numbers of strong lenses in upcoming surveys will be
significantly higher than predicted given a simple two-component model
that does not include AC. Second, if AC only occurs in massive,
early-type galaxies with particular shapes or formation histories,
then the ones for which AC does occur are more likely to be
strong-lensing systems.  Thus, for a given mass system, there may be a
selection bias favouring galaxies or groups depending on their
formation history.

Improved simulations, better understanding of galaxy formation, and
observational results that confirm, contradict, or refine these AC
models will be useful for understanding selection biases due to AC. A
more detailed exploration is beyond the scope of this paper.

\section{Concentration}
\label{S:concentration}

We now allow more model freedom than in the previous section, varying
both the dark matter inner slope $\gdm$ and the dark matter halo concentration
$\cdm$ on a grid (spherical model only) in
Section~\ref{SS:resultsconcgamma}.  We also consider a more limited set of
changes in dark matter halo concentration in the triaxial case in
Section~\ref{SS:resultsconctriax}.

\subsection{Concentration versus inner slope}
\label{SS:resultsconcgamma}

Here we explore the trade-off between steepening the dark matter inner
slope versus increasing the concentration (with fixed stellar component), both of which are expected
to increase the lensing cross-section due to the enhanced density of
dark matter in the very inner parts of the galaxy.

First, we explore this degeneracy with a relatively fine grid in dark
matter inner slope $\gdm$ and concentration $\cdm$, for spherical
cusped models only. Our grid has 10 values of $\gdm$ and 8 values of
$\cdm$ for each mass scale. In accordance with our finding in
Section~\ref{SS:resultsgamma} that the increase in total cross-section with
$\gdm$ for slopes steeper than unity is extremely rapid at our
fiducial concentration, we choose 10 values of $\gdm$ from 0 to 2,
distributed evenly in $\exp(\gdm)$. To define the grid in
concentration, we use the result from $N$-body simulations
that the distribution of $\cdm$ values is roughly lognormal around the
fiducial concentration, with scatter of $0.15$ dex
\citep{2001MNRAS.321..559B}. Our 8 values of $\cdm$ range from
$-3\sigma$ to $+4\sigma$, i.e., from $\cdm=3.0$ to $\cdm=33.4$ for the
lower mass scale, and from $\cdm=2.0$ to $\cdm=22.3$ for the higher
mass scale, in steps of $1\sigma$. (The grid is skewed towards higher values
because of our expectation that the cross-section will get quite large
in those cases, a trend we want to track). When we change $\gdm$
and/or $\cdm$, we normalise the profile by fixing $\Mv$ and allowing
$\rho_0$ to vary, as discussed in Paper~I.

For some of the grid points at high $\gdm$ and/or $\cdm$, the Einstein
radius is significantly larger than for our test case (with $\gdm=1.5$
and fiducial concentration) that we used to determine adequacy of the
box size in Paper~I.  To avoid box size effects for those grid points
we increase the box size by a factor of 3/2 (lower mass scale) or 2
(higher mass scale), while maintaining the pixel size. We expect any
remaining box size effects to reduce the cross-section for the
steepest slopes and highest concentrations, so any trends we see with
these parameters could be slightly underestimated.
For grid points at lower $\gdm$ and/or $\cdm$ than our
test case, the Einstein radius is smaller than the $\gdm=0.5$,
fiducial concentration case for which we tested resolution. However,
we find that it is at most 15 per cent smaller, because of the
unchanging stellar component which means the combined profile cannot
get arbitrarily small, and since our resolution choice was
conservative in the spherical case, we do not increase the resolution
any further.

\begin{figure*}
\begin{center}
$\begin{array}{c@{\hspace{0.2in}}c}
\includegraphics[width=3.0in,trim=0.5in 0 0 0]{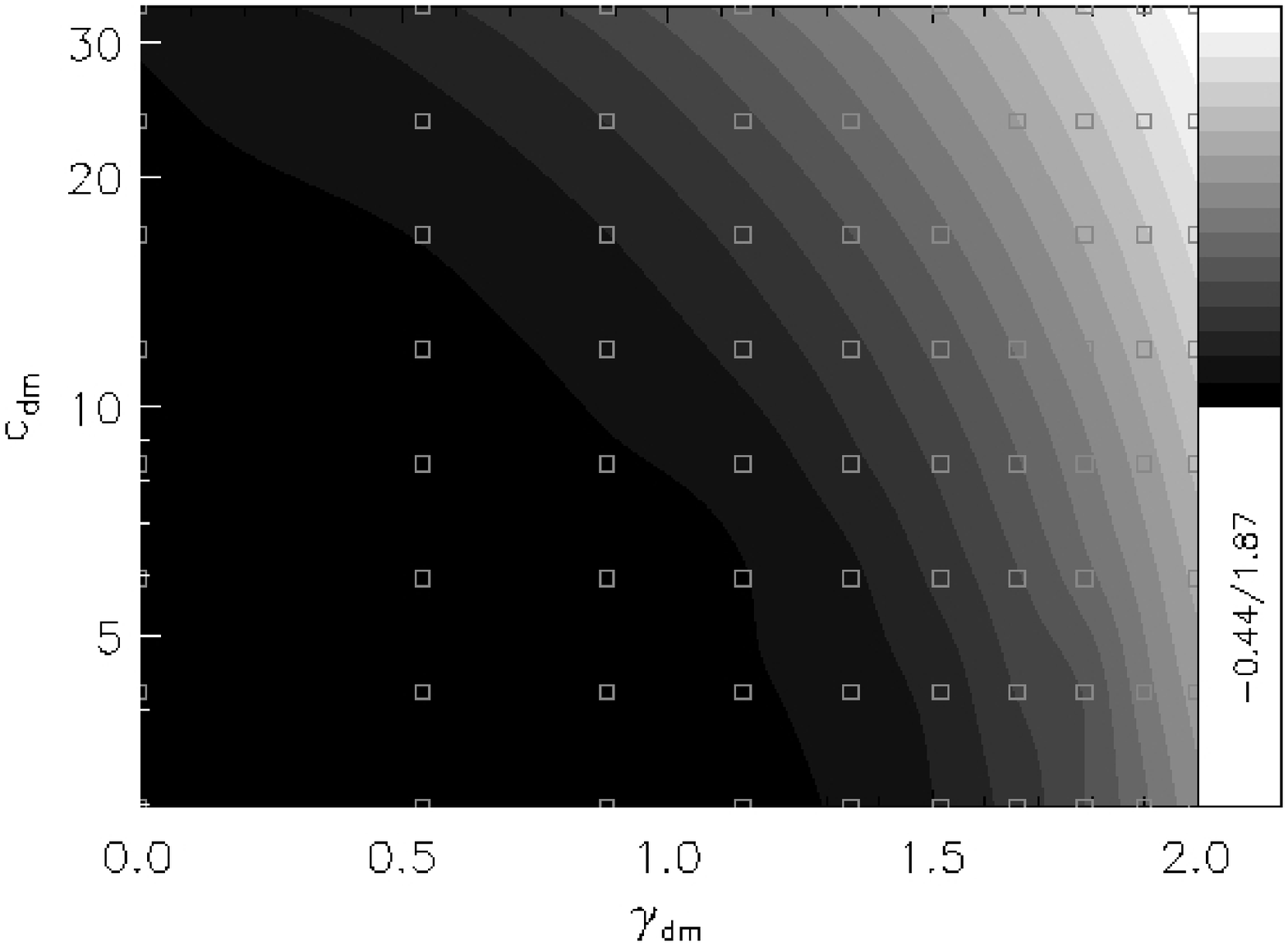} &
\includegraphics[width=3.0in,trim=0.5in 0 0 0]{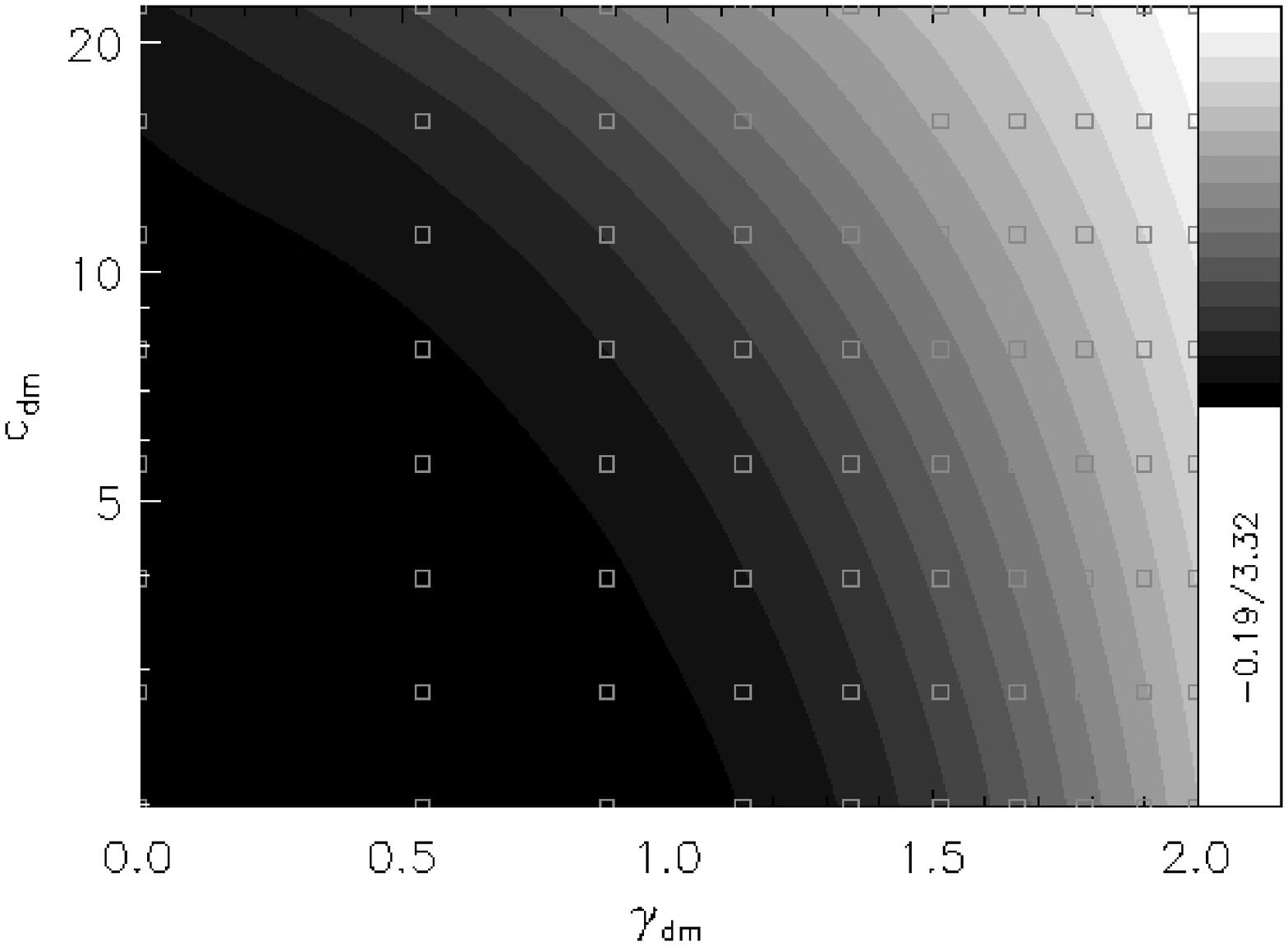} \\
\end{array}$
\caption{\label{F:gcgrid-fine} Contour plots of the 
  base-10 logarithm of the unbiased cross-section for the lower (left)
  and higher (right)  mass 
 models.   Note the different scales for the contour plots.  The
  points on the grid are shown with open squares, with the contours
  determined via interpolation.}
\end{center}
\end{figure*}

For each of the 80 grid points, we compute the unbiased and biased
lensing cross-sections; since the models are spherical, there is only
one cross-section to consider ($\st=\sig{2}$, $\sig{3}=\sig{4}=0$).
Fig.~\ref{F:gcgrid-fine} shows contour plots of the base-10 logarithm
of the lensing cross-section for both mass scales. For illustration
purposes, we show the unbiased cross-section only, and describe the
effects on the biased cross-sections in the text. For the lower mass
scale (left panel), the total variation of the unbiased cross-section
is a factor of 200; for the higher mass scale (right panel), a factor
of 3\,200. This difference in dynamic range for the two mass scales is
because the higher mass scale was already more dark matter-dominated
than the lower mass scale even for $\gdm=1$ and the fiducial
concentration, so increasing either one or both of those values has a
more dramatic effect on the cross-section. Much of this strong
variation with cross-section occurs in a fairly small range of $\gdm
\ge 1.3$, and the very highest cross-sections are only achieved for
concentrations that are several $\sigma$ above the median.
Nonetheless, the enormous variation in cross-section across the
parameter space suggests that the observed distribution of
concentration and inner slopes from strong lensing systems,
$p_\mathrm{SL}(\gdm,\cdm)$, may be quite biased compared
with the intrinsic distribution (at fixed mass).  Indeed, in
Section~\ref{SS:exampleapp} below we construct an example to
illustrate the bias in the distribution of \gdm\ and \cdm\ that
may be observed in strong lensing surveys.  We find that the
distribution can be both shifted and broadened by the lensing
selection bias.  These effects produce a significant (factor of
several) enhancement in the abundance of high-concentration halos that, according
to the intrinsic distribution, should constitute only $\sim 1$
per cent of the population.

Next, we look at slices through the grid. For the fiducial
concentration itself, the total variation in the cross-section with
$\gdm$ is a factor of 40 (lower mass scale) or nearly 600 (higher mass
scale) going from $\gdm=0$ to $\gdm=2$. For the fiducial inner slope
($\gdm=1$), as the concentration varies from $-3\sigma$ to $+4\sigma$,
the cross-section varies by a factor of five (lower mass) to several
tens (higher mass). Thus, we can conclude that (a) the effects of
varying concentration and inner slope of the dark matter are
significant, and (b) those effects are (not surprisingly) stronger in
systems where the dark matter is more dominant. In either case, it
seems clear that efforts to derive the underlying distribution of
$\gdm$ and $\cdm$ values using strong lensing-selected systems must
account for this strong bias in the cross-sections with dark matter
inner slope and concentration.

We also find that the trends of the cross-section with $\gdm$ and 
$\cdm$ do not depend too significantly on the magnification bias
modes. The total variation across the grid is somewhat smaller for the
biased cases than for the unbiased case, because for the steeper slope
models the cross-section does not increase as sharply as in the
unbiased case.

\begin{figure*}
\begin{center}
\includegraphics[width=0.9\textwidth]{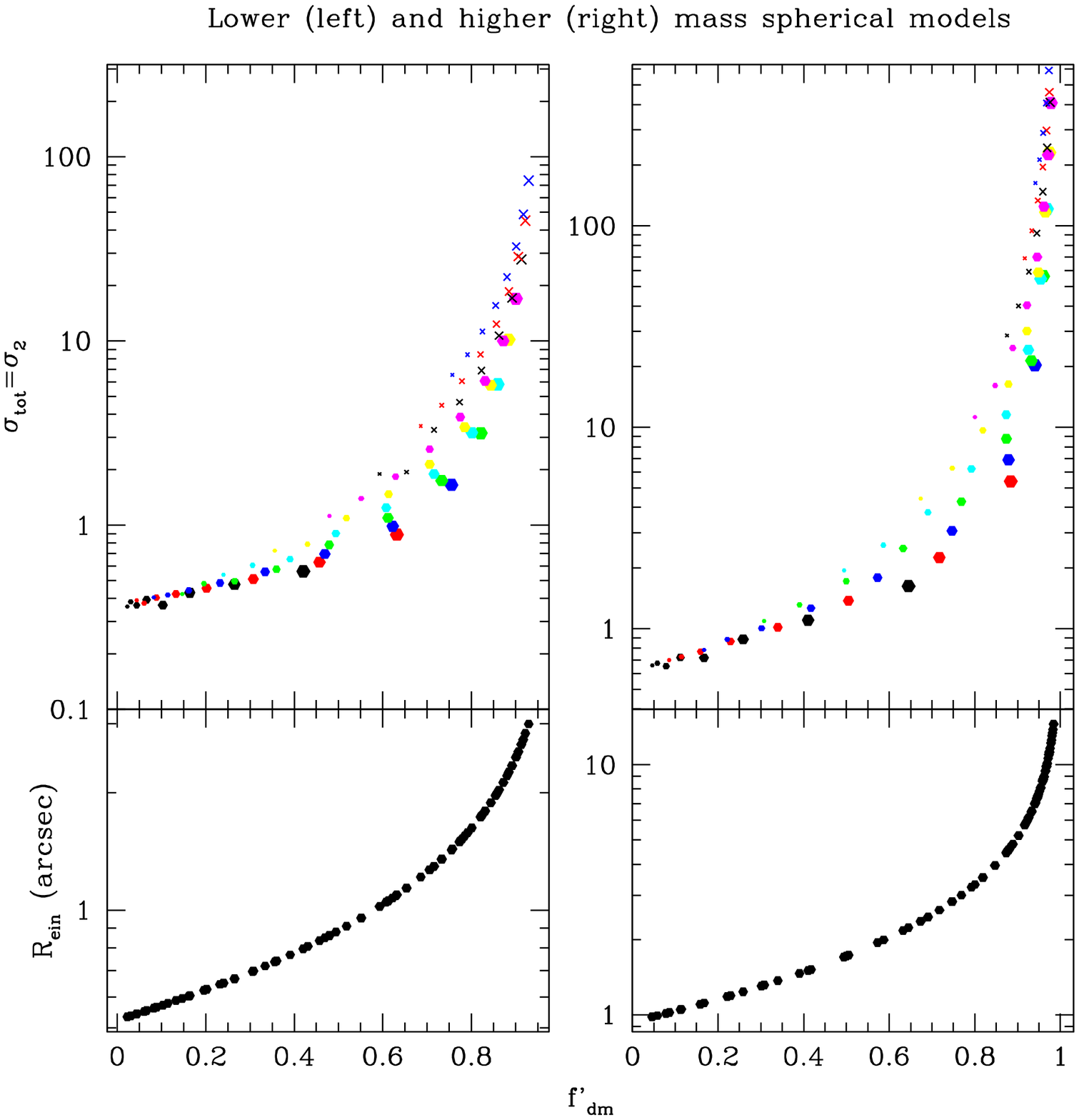} 
\caption{\label{F:fdm} For lower (left column) and higher (right
  column) mass scales, we show: (bottom)---The Einstein radius as a
  function of the projected dark matter fraction (within the Einstein
  radius) for all 80 models on the $\gdm$-$\cdm$ grid; and (top)---The
  cross-section as a function of the dark matter fraction.  The
  different point colours and types are the values at fixed $\gdm$:
  hexagons in $\{$black, red, 
  blue, green, cyan, yellow, magenta$\}$ for $\gdm=\{0.00, 0.54, 0.88,
  1.14, 1.35, 1.52, 1.66\}$, and crosses in $\{$black, red, blue$\}$
  for $\gdm=\{1.79, 1.90, 2.00\}$.  The symbol sizes from smallest to
  largest indicate the $8$ increasing values of $\cdm$ at fixed
  $\gdm$.}
\end{center}
\end{figure*}

We further explore the degeneracy between $\gdm$ and $\cdm$ by
considering, for all models on the grid, the Einstein radius $\Rein$
and the (projected) dark matter fraction $\fpdm$ within $\Rein$. We
expect both to be connected to the lensing cross-section $\st$ in some
non-trivial way.  Fig.~\ref{F:fdm} shows the connection between all
three quantities for both mass scales (unbiased cross-section only).

First, the relation between the Einstein radius and the dark matter
fraction shown in the bottom panels is perfectly one-to-one. This fact
follows from our variation of the dark matter profile while fixing the
stellar component (see Appendix~\ref{A:dmfrac} for a derivation of
this result). Second, the relation between the lensing cross-section
and the dark matter fraction, shown in the top panels, is \emph{not}
one-to-one, though there clearly is a mean relation with some scatter
around it. The Einstein radius is computed in the lens plane, while
the cross-section is computed in the source plane. The mapping between
the two planes depends on the density profile in a different way than
the dark matter fraction depends on the density profile. Therefore,
variations in $\cdm$ and $\gdm$ affect the mapping, and thus $\st$, in
a different way than they affect $\fpdm$, resulting in the observed
scatter.

While the mass range under consideration is somewhat different, we can
compare our higher mass model qualitatively against the results in
\cite{2007A&A...473..715F} for the effect of halo concentration on the
strong lensing cross-section.  In that paper, they use
single-component NFW halos from simulations to model lensing by galaxy
clusters (for which this single-component model is not as bad an
approximation as it is for galaxy-scale halos, where the baryonic
component must be incorporated to obtain even remotely plausible
lensing properties).  Our results are qualitatively similar to the
results in \cite{2007A&A...473..715F}, who find that: (a) for a sample
with a broad mass range, the lensing cross-section is dominated by
those of the highest mass and therefore lowest concentration; and (b)
for a sample with a narrow mass range, it is dominated by those
clusters at the higher end of the lognormal scatter in halo
concentration at fixed mass.  Their results are consistent with our
finding that the mass increase going from lower to higher mass scale
has a far stronger effect on $\st$ than the decrease in concentration,
but that at fixed mass, changes in concentration can alter the lensing
probabilities by $\sim 50$ per cent or more.  Thus, the abundance of
high-concentration halos as lenses should not be taken as a reflection
on the underlying distribution of concentrations; and likewise, claims
that too many high-concentration clusters have been observed should be
revisited with a concentration distribution from simulations modified
by $\st$ as a function of mass and concentration.

We also compare against the results of \cite{2004ApJ...601..104K}, who
use spherical, single-component NFW model to understand the lensing
properties of cluster DM halos. That paper finds results similar
to ours: first, that increases in concentration increase $\st$;
second, that these increases correlate with the mean of the image
separation distribution (and therefore $\Rein$), so the distribution
of concentration parameters will impact the observed image separation
distribution for lensing systems.  We can compare our
Fig.~\ref{F:inversion} (lower left) against their fig.~6, which shows the median
of the concentration distribution shifting by approximately 70 per
cent when accounting for lensing selection bias. Our shift is
significantly lower than that ($14$ per cent) because their analysis uses NFW halos
only whereas ours includes a non-negligible stellar component that
will tend to reduce the effects of changing the NFW concentration.
The different mass range may also play a role in the difference.

These results have implications for the results in
\cite{2007MNRAS.379..190C}, which includes a 
compilation of inferred NFW profile concentrations for strong
lensing-selected clusters and a fit to a power-law
concentration-mass relation.  Ignoring issues of possible modeling
bias (for example, due to overly simple modeling of cluster shape
or substructure), we can
attempt to explain the fact that the normalisation of their
concentration-mass relation is at least 20 per cent higher than
in $N$-body simulations.  While they consider the possibility that
this high normalisation may be due to adiabatic contraction, we
note that for our group-scale model (as in Fig.~\ref{F:inversion})
the median of the concentration distribution at fixed halo mass
was increased due to selection bias by nearly that amount.  There is of
course the caveat that their mass range is somewhat higher than ours,
but the results of \cite{2007A&A...473..715F} also support this
possibility.  Consequently, adiabatic contraction (which has not been
proven to occur in cluster-mass halos) may not be needed, and
selection bias may be significant enough to 
explain nearly all of this apparent discrepancy with simulations.

\subsection{Varying concentration in the triaxial case}
\label{SS:resultsconctriax}

We explore the variation of orientation-averaged cross-sections with dark matter
concentration more coarsely for triaxial models, to verify that the
trends remain roughly the same as for spherical models. This
assumption is acceptable for our coarse grid in $\gdm$ as presented in
Section~\ref{SS:resultsgamma}, but we need to confirm this for $\cdm$ as
well. Because of the need for 24 viewing angles for any triaxial
model, we use only one value of $\gdm=1$, and we restrict the grid in
$\cdm$ to five values ranging from $-2\sigma$ to $+2\sigma$ in steps
of $1\sigma$.  Thus for the lower mass scale, we go from $\cdm=4.2$ to
$\cdm=16.8$; for the higher mass scale, from $\cdm=2.8$ to
$\cdm=11.2$. The actual values at the fiducial concentration are
given in Table~\ref{T:haloshape} for all magnification bias modes, so
we only present trends relative to those values.

\begin{figure}
\begin{center}
\includegraphics[width=\columnwidth,angle=0]{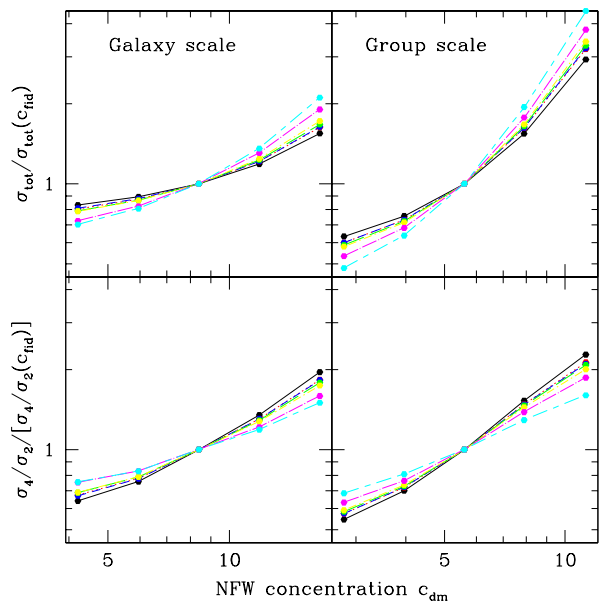} 
\caption{\label{F:varyc}For lower (left column) and higher (right
  column) triaxial mass models, we show the variation of the total,
  orientation-averaged lensing cross-section (normalised to the value
  at the fiducial concentration) with concentration in the top panels,
  and the variation of the 4:2 image ratio with concentration in the
  bottom panels.  Line colours and types indicate the magnification
  bias mode, as described in the caption for Fig.~\ref{F:incoblpro}.}
\end{center}
\end{figure}

There are several points we infer from the resulting
Fig.~\ref{F:varyc}. First, our conclusions from the spherical case
are, broadly speaking, robust for the triaxial case: variation of
$\cdm$ with all other parameters fixed leads to enhancements of the
total lensing cross-section by factors of a few.  In particular, the
enhancement can be factors of $\sim 2$ for the lower mass scale, and
$\sim 4$ for the higher mass scale, where the dark matter component is
more important. The reason the range of enhancements is smaller than
in Section~\ref{SS:resultsconcgamma} is that the range in $\cdm$ is
smaller; however, the range of enhancements is in line with the
spherical case when we restrict the grid for the calculations with
spherical models in the same way.

One point we were not able to test using the spherical case is the
issue of the quad/double ratio $\sig{4}/\sig{2}$. Here, we see that
this number also can change by factors of a few for reasonable
ranges of concentrations for both mass scales.  This change presumably
results from the fact that the dark matter is intrinsically less round
than the stellar component, so increasing the dark matter
concentration increases its importance in the inner parts and
therefore increases the projected ellipticity of the total mass
distribution.

We can also contrast Fig.~\ref{F:varyc} with the ``triaxial'' panels
of Figs~\ref{F:gammacs} and~\ref{F:gamma42}, showing the variation of
respectively $\st$ and $\sig{4}/\sig{2}$ with $\gdm$ at the fiducial
concentration. First, the increase in total lensing cross-section when
going from the median concentration to that $+2\sigma$ is about half
the increase when going from $\gdm=1$ to $\gdm=1.5$. Second, there are
some subtleties that mean that increasing $\gdm$ and increasing $\cdm$
do not affect the lensing cross-section in quite the same way. In
particular, the ranking of magnification bias modes is different in
the two cases: when increasing $\gdm$, the $\st$ for the higher flux
limit magnification bias modes does not increase as much as for the
unbiased case, whereas the opposite is true when increasing $\cdm$.
Furthermore, $\sig{4}/\sig{2}$ tends to flatten out when going from
$\gdm=1$ to $\gdm=1.5$, whereas it increases by a factor of $\sim 2$
when going from the median concentration to $+2\sigma$. Both of these
differences between increasing $\gdm$ and increasing concentration are
due to the fact that the biased cross-sections and the four-image
cross-section are related not just to the size of the region in which
lensing can occur, but also to the scaling of the surface mass density
with radius within this region.

Despite some subtle differences between changing concentration and
inner slope, it is clear that in strong lensing-selected analyses, we
expect significant selection biases (factors of several) favouring
steeper inner slopes $\gdm$ and higher concentrations $\cdm$ at fixed
mass.  This selection bias is manifested not just in $\st$ but also in
$\sig{4}/\sig{2}$. This result implies that an analysis that recovers
the distribution of inner slope and/or concentration with high
statistical precision does not measure the intrinsic distribution of
these quantities directly, but must account for the selection biases
first using simulations.

\section{Deprojected S\'ersic density}\label{S:sersic}

In this section, we investigate selection bias with density profile
parameters for the composite deprojected S\'ersic density profiles.
Previous work has characterised the basic lensing properties of
spherical, single-component deprojected S\'ersic models
\citep{2004A&A...415..839C,2009JCAP...01..015B}, but we study the
full lens statistics of two-component spherical and non-spherical
models.  We divide this analysis into three parts, where we first
address the comparison against cusped models in the spherical
case; we next consider orientation dependence in the mixed and
triaxial cases; and we finally allow the S\'ersic index to vary
separately for the dark matter and stellar components.

\subsection{Comparison against cusped models}

We first ask whether deprojected S\'ersic models lead to significantly
different lensing cross-sections compared to cusped models, for both
mass scales in the spherical case. This question is important for
example in theoretical calculations of the expected number of lenses
from a particular strong lensing survey.

The fiducial cusped models with $\gdm=\{0.5,1.0,1.5\}$ have projected
dark matter fractions within the corresponding Einstein radii of
$\fpdm=\{0.11, 0.27, 0.62\}$ (lower mass) and $\fpdm=\{0.18, 0.49,
0.88\}$ (higher mass).  The fiducial deprojected S\'ersic model has
$\fpdm=0.55$ and $\fpdm=0.85$ for lower and higher mass scales,
respectively, which means that qualitatively we expect the Einstein
radii and cross-sections to be more comparable to the $\gdm=1.5$
cusped model than to the $\gdm=1$ case.  The results from analytic
\gravlens\ calculations show that this expectation is correct: for the
lower mass scale, $\st=1.57$\,arcsec$^2$, comparable to
$\st=1.40$\,arcsec$^2$ for the $\gdm=1.5$ case, while
$\st=0.53$\,arcsec$^2$ for $\gdm=1$. Likewise, for the higher mass
scale, $\st=10.5$\,arcsec$^2$, which is closer to
$\st=15.9$\,arcsec$^2$ for the $\gdm=1.5$ case than to
$\st=1.63$\,arcsec$^2$ for the $\gdm=1$ case. All numbers quoted here
are unbiased cross-sections, and the trends for $\Rein$ are similar.
We conclude that for the same mass within spheres of over-density
$180\overline{\rho}$, deprojected S\'ersic models with
observationally-motivated S\'ersic indices and half-mass radii tend to
give higher lensing cross-sections than cusped models with the
``canonical'' inner slope value $\gdm=1$.  It is clear that the
selection bias with mass for the deprojected S\'ersic models is
broadly consistent with what one would expect for the cusped models
with $\gdm=1.5$, which is stronger than the selection bias for the
$\gdm=1$ models.

When we include the effects of magnification bias, we find that the
enhancement in the cross-section is similar to the enhancement for the
$\gdm=1.5$ cusped models.  The image separation distributions are
fairly narrow as for the cusped models, with a mean of $1.98\,\Rein$
(lower mass) and $1.88\,\Rein$ (higher mass).  This result is
consistent with our finding in Paper~I that the composite deprojected
S\'ersic model for the galaxy scale is very close to isothermal,
whereas the higher mass scale is more dominated by the dark matter and
therefore is somewhat less isothermal.

\subsection{Orientation-dependence}\label{SS:sersicorient}

We had found, in Section~\ref{SS:resultsorient}, that for the triaxial and
mixed shape cusped models, the total unbiased cross-section $\st$ can
vary by factors of several when we change the viewing angles.
Likewise, the ratio 
$\sig{4}/\sig{2}$ can vary from near zero to $\sim 15$
per cent.

We now compare this result for the orientation dependence in cusped
models against the results for the deprojected S\'ersic models.  We
find that the results are very similar for variation of both $\st$ and
of $\sig{4}/\sig{2}$, which suggests that these numbers do not depend
strongly on the functional form of the density profile (for reasonable
models).  Furthermore, when we include the effects of magnification
bias, we find that it changes \st\ and $\sig{4}/\sig{2}$ in similar
ways as it does for the cusped, $\gdm=1.5$ models.  This change includes
sharp increases in \st\ and significant enhancement of the quad/double ratio.

When averaging over orientation, we also find that, as for cusped
models, the value of \st\ does not depend strongly on halo shape (when
comparing triaxial, mixed, and spherical models).

\subsection{S\'ersic index}
\label{SS:sersicindex}

In this section, we consider varying the S\'ersic indices of both the
dark matter ($\ndm$) and stellar component ($\ns$), in the spherical
case, for both mass scales. All variation is done at fixed mass.
Because an increase in S\'ersic index leads to a more highly
concentrated profile in the inner regions, we expect that with
appropriate normalisation (discussed in Paper~I), the lensing
cross-section will be an increasing function of S\'ersic index.  The
questions we address are (1) the extent of the selection bias that is
possible for a reasonable range of S\'ersic indices, and (2) the
relative importance of the S\'ersic index for each component of the
profile.  We adopt the ranges in $\ndm$ and $\ns$ as derived in
Paper~I, and sample five values equally in both $\log\ns$ and
$\log\ndm$, resulting in a $5 \times 5$ grid.  For $\ndm$ and $\ns$
values different from the fiducial values (in the centre of the grid),
we change $\rho_0$ to maintain the total mass and scale radius.

\begin{figure}
\begin{center}
$\begin{array}{c}
\includegraphics[width=\columnwidth,angle=0]{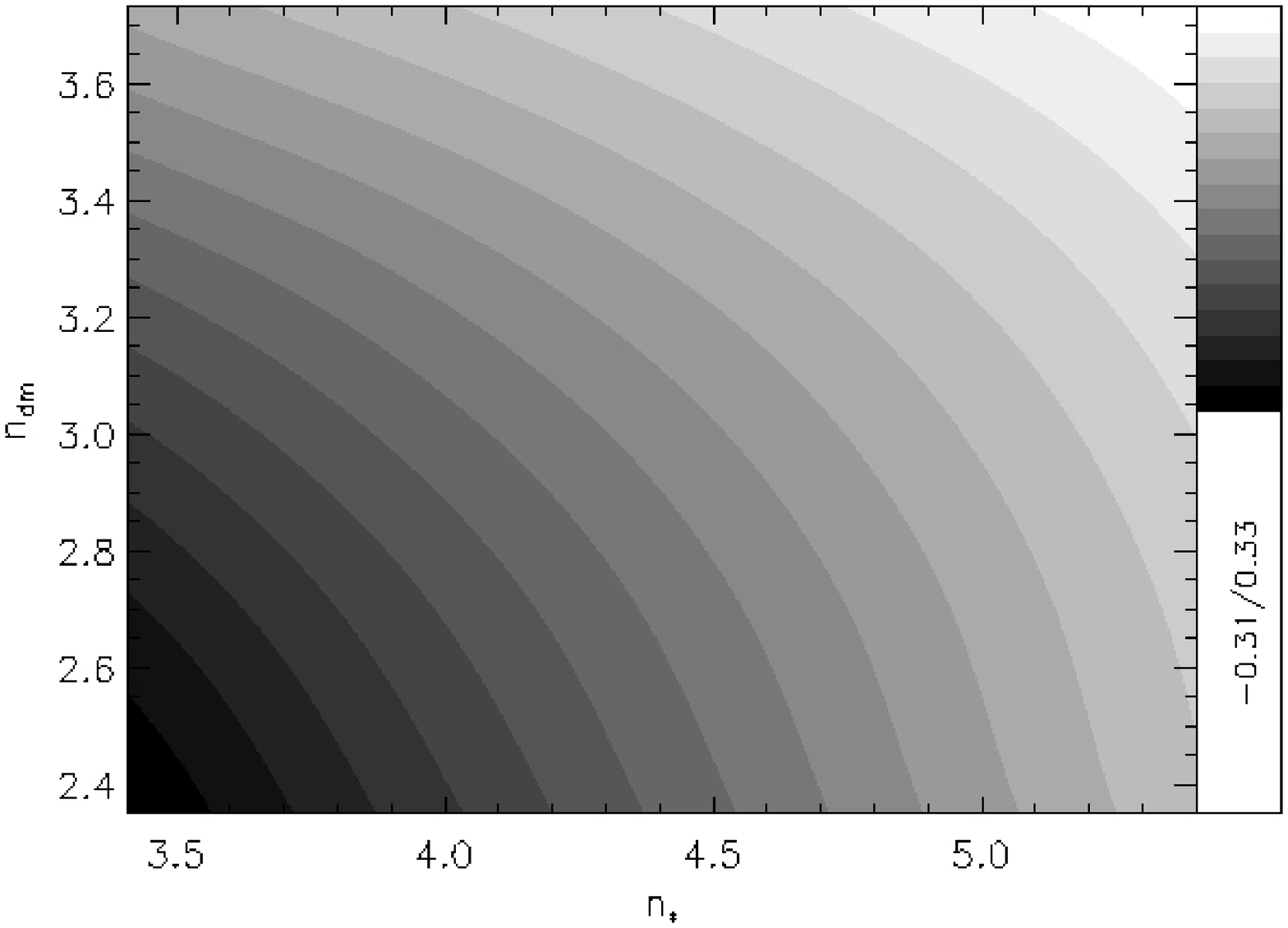} \\
\includegraphics[width=\columnwidth,angle=0]{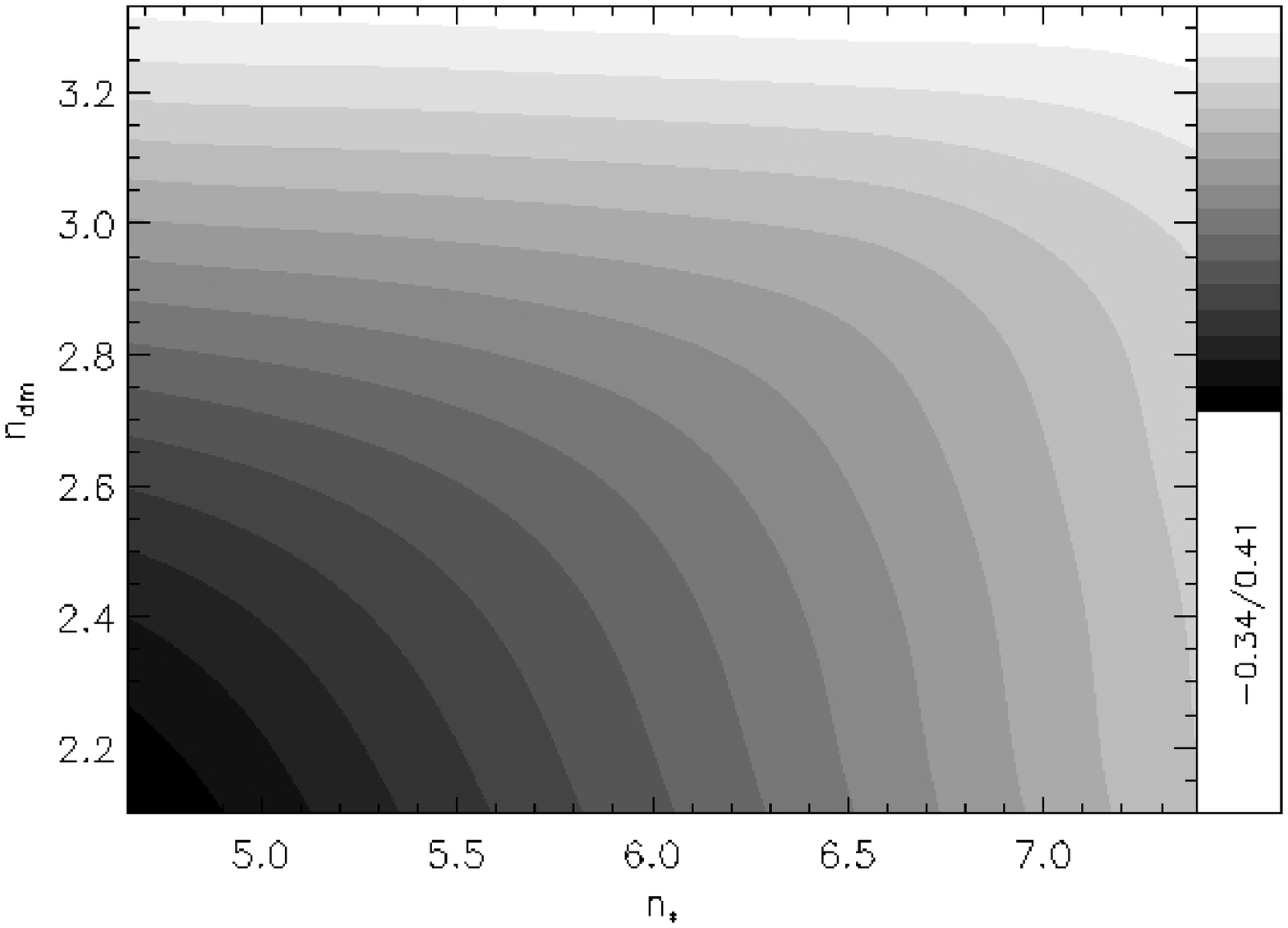} \\
\end{array}$
\caption{\label{F:ngrid}Contour plots of the base-10 logarithm of the
  unbiased cross-section for the lower (top) and higher (bottom) mass
  scales, relative to the cross-section at the
  fiducial grid point. }
\end{center}
\end{figure}

Fig.~\ref{F:ngrid} shows the unbiased cross-section as a function of
the S\'ersic indices for both mass scales, with contours in the base-10
logarithm relative to the cross-section at the fiducial grid point,
i.e., $\log{(\st/\stfid)}$. The increase in cross-section with both
$\ndm$ and $\ns$ is physically understandable.  As shown, the total
variation on the grid is from $\sim 0.5$ to $2\,\stfid$ for the lower
mass scale (factor of 4), or $\sim 0.45$ to $2.6\,\stfid$ for the
higher mass scale (factor of 6). While these increases are
significant, they are more comparable to the changes in cross-section
for the cusped models when varying $\cdm$ than varying $\gdm$.  The
reason for this is that changing $\gdm$ for a cusped model causes a
large change in the inner parts of the intrinsic and projected
profile, but changing the S\'ersic index leads to substantial changes in both the
inner and the outer intrinsic profile that partially cancel out when
projected along the line-of-sight.  The shape of the contours, which
run diagonally at low $\ndm$ and $\ns$, but closer to horizontally at
high $\ndm$ and vertically at high $\ns$, is also physically
understandable: at low $\ndm$ and $\ns$, both components are roughly
equally important, whereas at high $\ndm$, the stellar component (and
therefore $\ns$) is less important, and at high $\ns$, the dark matter
component (and therefore $\ndm$) is less important.  The effect at
high $\ndm$ is more exaggerated for the higher mass scale, because
this model was designed to be more dark-matter dominated than the
lower mass scale model.

\begin{figure}
\begin{center}
\includegraphics[width=\columnwidth,angle=0]{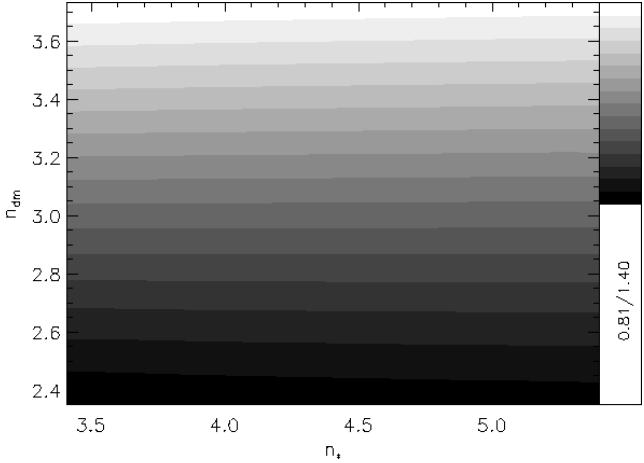} 
\caption{\label{F:reinoverrh} Contour plot of $\Rein/R_{h,*}$ for the
  lower mass deprojected S\'ersic models. }
\end{center}
\end{figure}

\begin{figure}
\begin{center}
\includegraphics[width=\columnwidth,angle=0]{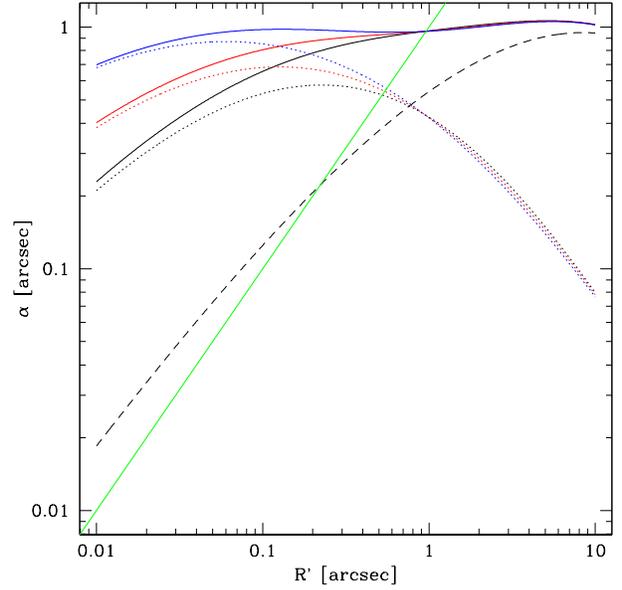} 
\caption{\label{F:sersicdef} Deflection as a function of scale for the
  lower mass deprojected S\'ersic mass model. The black dashed line is
  for the fiducial DM component; the black, red, and blue dotted lines
  are for the stellar components with the lowest, fiducial, and
  highest $\ns$; and the black, red, and blue solid lines are the
  corresponding composite models (stellar plus DM).  Finally, the
  green solid line corresponds to $\alpha=R'$; the intersection of
  this line with the deflection curves occurs at $\Rein$.}
\end{center}
\end{figure}

Fig.~\ref{F:reinoverrh} shows the trends of $\Rein/\Rhs$ for the lower
mass scale on this grid.  Note that since we normalise the profiles by
fixing the scale radius and mass while varying $\rho_0$, the scale
radius $\Rhs$ is the same at each point on the grid, so only $\Rein$
is varying. The exceptionally horizontal contours require some
explanation, since the cross-section contours in Fig.~\ref{F:ngrid}
are not strictly horizontal.  This result can be explained by the
deflection curves, shown in Fig.~\ref{F:sersicdef} for this mass
scale.  In this plot, the Einstein radius is where the one-to-one line
crosses the $\alpha(R')$ curves for the composite models, and as we
have already seen, these crossings occur in nearly the same place.
This results from our normalisation convention, which fixes the same
half-mass radius $\Rhs$ for each of the $\ns$ values, meaning that it
fixes the average convergence $\overline{\kappa}(\Rhs)$ and therefore
the deflection at $\Rhs$, $\alpha = \Rhs \overline{\kappa}(\Rhs)$, to
the same value.  Because the DM component is also fixed, it means that
the models with different $\ns$ have the same $\Rein$ and the same
$\fpdm$ within $\Rein$. However, this is purely a consequence of our
chosen redshifts and the balance of stellar and dark matter in our
models; had we arranged it so that $\Rein \ll \Rhs$, then $\ns$ would
indeed affect $\Rein$.
%
However, the cross-sections do not have to be the same despite
identical $\Rein$ values for varying $\ns$, because they are set by
the outer (source-plane) caustic, which is related to the inner
(lens-plane) critical curve, determined by the \emph{slope} of the
deflection on small scales. Clearly, in Fig.~\ref{F:sersicdef}, while
the deflections are the same at $\Rein$, their slopes are quite
different within $\Rein$, leading to different lensing cross-sections.
Finally, we note that despite our superimposing two non-trivially
complicated deprojected S\'ersic models to make the composite galaxy
models, the deflection curve for the composite model is remarkably
close to the flat, isothermal case ($\alpha = \Rein$), which we have
also seen with the cusped models. This fact explains the very narrow
image separation distribution for all of these composite models.

\begin{figure}
\begin{center}
  \includegraphics[width=\columnwidth,angle=0]{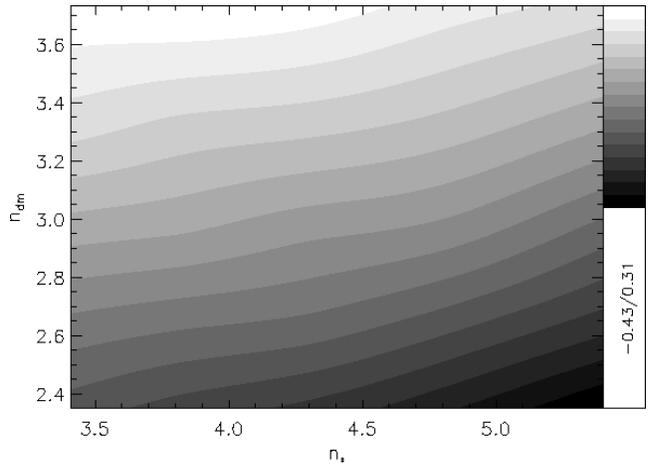}
\caption{\label{F:ngrid-biased} Contour plot of the base-10 log of
  the biased cross-section for the lower mass scale, relative to this
  cross-section at the fiducial grid point. }
\end{center}
\end{figure}

Finally, we consider the selection bias trends for biased
cross-sections, particularly the one using the highest flux limit and
total magnification. That trend is presented for the lower mass scale in
Fig.~\ref{F:ngrid-biased}, and shows that the shape of the contours is
quite different from the unbiased case (Fig.~\ref{F:ngrid}, top
panel).  While \ndm\ has the same, significant effect on the
cross-section as in the unbiased case, the effect is much less
significant for \ns\ and has changed sign: stellar components that are
less concentrated give slightly higher biased cross-sections.  While
there is no simple, physical explanation for this result, it is
apparently related to the scaling of the convergence within the
Einstein radius, and the fact that a very concentrated (high \ns)
stellar component only has a very small region where the convergence
is extremely high, whereas a less concentrated one has a larger area
where it is fairly high, which turns out to be more important for
these very biased cross-sections.

\section{Robustness of conclusions to changes in redshift}\label{S:robustness}

Finally, to ascertain how much our conclusions about selection biases
in the previous sections depend on our chosen fiducial lens and source
redshifts, we now try varying the redshifts in several ways.

If we vary both the lens and the source redshifts in a way that
preserves $\Sigma_c$, then the physical scale within the galaxy that
is measured by strong lensing is the same, but it will lead to
different image separation distributions, scaled by the angular
diameter distance $D_L$ of the lens galaxy. This change may lead to
{\em observational} selection effects that depend sensitively on the
survey image search strategy (resolution and maximum angular
separation), but no physical selection effects, so we do not attempt
to model this in any general way.

Now consider fixing the lens redshift (and therefore the conversion
from physical to angular scale), but shifting the source redshift to
rescale the critical surface mass density $\Sigma_c$. This shift will
change the Einstein radius $\Rein$ and therefore the physical scale
of the density profile that is probed by strong lensing.  Thus,
this change may result in different \emph{physical} selection biases.

If we fix the lens redshift to our fiducial $z_L=0.3$ but move the
source redshift considerably lower, from the fiducial $z_S=2.0$ to
$z_S'=0.6$, then the critical surface density increases by $\sim 70$
per cent. This change means that the strong lensing is due to smaller
scales, where the average projected surface density is higher, and that
the lensing cross-section for this density profile will be lower.
Because of the importance of smaller scales, the projected dark
matter fraction $\fpdm$ will decrease. As a test, we recompute \Rein\
and the lensing cross-sections for both mass scales as a function of
the dark matter halo inner slope $\gdm$ 
to see the degree to which the selection bias we identified with our
fiducial lens and source redshifts persist. We also increase the
resolution of the surface density maps used by \gravlens\ for these
calculations by a factor of two, while maintaining the box size, to
ensure that the critical curves can be properly resolved.

We also consider a value of $\Sigma_c$ that is 40 per cent lower than
our fiducial value.  While this low $\Sigma_c$ does not correspond to
any sensible value of $z_S$ for our fiducial $z_L$, we assume a higher
$z_L\sim 0.6$ and $z_S\sim 3$.  Because of the different $z_L$, all
angular scales output by our pipeline must be rescaled by
$D_A(z=0.3)/D_A(z=0.6)=0.66$, i.e., the Einstein radii must be
rescaled by $0.66$ and the cross-sections by $0.66^2$.  However, the
relevant quantities for selection bias associated with true variation
of galaxy density profiles, e.g. $\fpdm$, are preserved.  Since we
expect increased Einstein radii and cross-sections, we preserve the
resolution but increase the box size by a factor of $2$.

Table~\ref{T:varyz} shows what happens to relevant quantities for
lensing when we vary $\Sigma_c$ to values 70 per cent higher and 40 per cent
lower.  We consider only the unbiased cross-sections for the spherical
shape model. 
\begin{table*}
\caption{\label{T:varyz}Summary of the lensing results for the cusped, spherical
  model when varying $\Sigma_c$ so that different physical
  scales are probed.  The factor of $0.66$ in \Rein\ for the lower
  $\Sigma_c$ cases is shown separately since it is not indicative of
  the physical scale within the galaxy.}
\begin{tabular}{l|r|r|r}
\hline
\hline
Quantity & Fiducial $\Sigma_c$ & Higher $\Sigma_c$ & Lower $\Sigma_c$ \\
\hline
\multicolumn{4}{c}{Lower mass scale} \\
$\Rein(\gdm=0.5)$          & $0.58$ & $0.36$ & $0.79\times 0.66=0.52$ \\
$\Rein(\gdm=1.0)$ (arcsec) & $0.68$ & $0.41$ & $0.95\times 0.66=0.63$ \\
$\Rein(\gdm=1.5)$          & $1.04$ & $0.61$ & $1.46\times 0.66=0.97$ \\
$\fpdm(R'<\Rein,\gdm=0.5)$ & $0.13$ & $0.08$ & $0.17$ \\
$\fpdm(R'<\Rein,\gdm=1.0)$ & $0.28$ & $0.20$ & $0.37$ \\
$\fpdm(R'<\Rein,\gdm=1.5)$ & $0.60$ & $0.51$ & $0.67$ \\
$\st(\gdm=0.5)/\st(\gdm=1.0)$ & $0.79$ & $0.79$ & $0.79$ \\
$\st(\gdm=1.5)/\st(\gdm=1.0)$ & $2.65$ & $2.96$ & $2.51$ \\
\hline
\multicolumn{4}{c}{Higher mass scale} \\
$\Rein(\gdm=0.5)$          & $1.18$ & $0.63$ & $1.73\times 0.66=1.14$ \\
$\Rein(\gdm=1.0)$ (arcsec) & $1.75$ & $0.88$ & $2.75\times 0.66=1.82$ \\
$\Rein(\gdm=1.5)$          & $4.44$ & $2.33$ & $6.75\times 0.66=4.46$ \\
$\fpdm(R'<\Rein,\gdm=0.5)$ & $0.22$ & $0.13$ & $0.30$ \\
$\fpdm(R'<\Rein,\gdm=1.0)$ & $0.51$ & $0.36$ & $0.63$ \\
$\fpdm(R'<\Rein,\gdm=1.5)$ & $0.87$ & $0.80$ & $0.91$ \\
$\st(\gdm=0.5)/\st(\gdm=1.0)$ & $0.52$ & $0.51$ & $0.52$ \\
$\st(\gdm=1.5)/\st(\gdm=1.0)$ & $9.19$ & $12.9$ & $8.04$ \\
\hline
\hline
\end{tabular}
\end{table*}
First, we can see that when we increase (decrease) $\Sigma_c$, we
decrease (increase) \Rein\ and \fpdm, as expected.  The change in
$\Rein$ can be significant, up to a factor of $\sim 2$ for the changes
in $\Sigma_c$ that we have tested.  We find that the selection bias
with $\gdm$, represented in the last two rows of this table, is nearly
independent of $\Sigma_c$ for $\gdm$ below the fiducial value of $1$,
and for higher $\gdm$ it is a slightly increasing function of
$\Sigma_c$.  The sign of this change is surprising, in the sense that
if \fpdm\ is lower than we might naively expect the parameters of the
dark matter distribution to be less significant. However, we can
explain this change by the fact that the cross-section \st\ is
determined by the outer (source-plane) caustic, which relates to
the inner (lens-plane) critical curve.  This in turn is determined
by the slope of the density profile on small scales, so if the
lensing properties are determined by smaller (larger) physical
scales, as when we increase (decrease) $\Sigma_c$, then \gdm\ becomes
more (less) important and selection bias becomes stronger (weaker).

\begin{figure}
  \begin{center}
    \includegraphics[width=\columnwidth,angle=0]{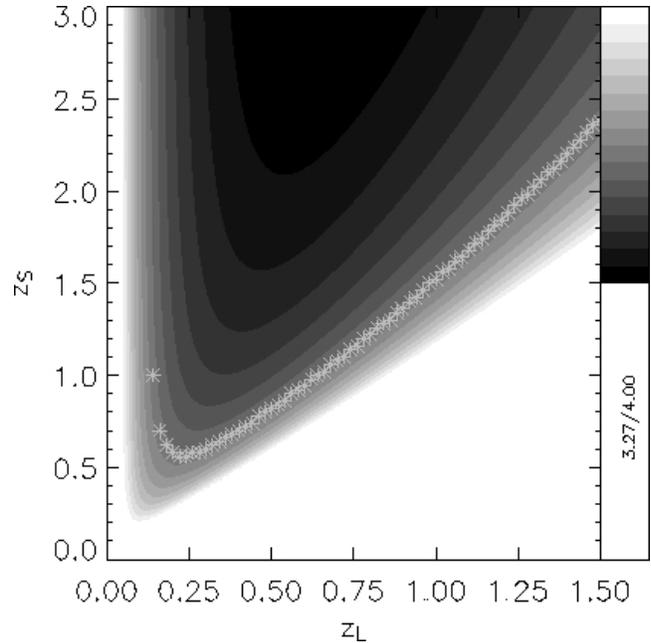}
    \caption{\label{F:sigmacrit}Critical surface density, plotted as
      $\log{[\Sigma_c/(M_{\odot}/\mathpc^2)]}$, 
        as a function of lens and source redshift.  For $z_S<z_L$,
        where $\Sigma_c$ is formally infinite, we have set it to the
        maximum value shown, $10^4 M_{\odot}/\mathpc^2$. Asterisks
        indicate the minimum source redshift for a given lens redshift
      for which we have confirmed that our conclusions about selection
      bias are valid.}
  \end{center}
\end{figure}

However, it is encouraging that the selection bias only changes by
$+14$ and $-4$ per cent ($+41$ and $-13$ per cent) for changes in
$\Sigma_c$ of $+70$ and $-40$ per cent for the lower (higher) mass
model.  This result suggests that within the range of $\Sigma_c$ that we have
considered in this subsection, the conclusions of our paper regarding
physical selection bias are not strongly dependent on our choice of
fiducial lens and source redshifts.  (The same is not true for
observational selection biases, which may be strongly affected by
changing $z_L$ and thus the characteristic angular scale of the
system.)  On Fig.~\ref{F:sigmacrit}, we have indicated the lens and
source redshift combinations for which our conclusions are robust
according to the tests in this subsection.

We must also consider the issue of magnification bias.
Magnification bias complicates a change in the redshift of the system
in two ways.  First, if the source redshift is different, then for a
given survey flux limit, the luminosity threshold is different.  For a
broken power-law luminosity function that is steeper at the bright
end, this means that if the source is more distant, the limiting
luminosity becomes brighter, so the actual number of lensing systems
is modulated due to both the different number density of sources above
that limit and by the change in effective $\st$ due to the steeper
luminosity function at the limit.  Second, since the source luminosity
function may be a function of redshift, we anticipate that the biased
cross-sections, which are the relevant quantity for any particular
survey given its flux limit and search strategy, may also change due
to this effect. For our analysis in this section, we have ignored this
change in the magnification 
bias formalism, but to determine the cross-sections for a real survey
this evolution would have to be taken into account.   In any case, both of these
magnification bias 
effects qualify as observational rather than physical selection biases
that are more properly simulated in the context of a particular strong
lensing survey.

\section{Conclusions}
\label{S:conclusions}

Using the pipeline developed in Paper~I, we have presented a series of
tests for selection biases in point-source strong lensing surveys due
to physical differences between galaxy or group mass systems.  These
tests reveal whether the distribution of physical parameters, such as
profiles and shapes of (projected) densities, in strong lensing systems
reflects that distribution among the galaxy population in general.

We have attempted to distinguish between physical selection biases
which relate to the parameters of the galaxy density, versus
observational selection biases which depend on the lensing detection
strategy (specifically, the flux limit, angular resolution, and
maximum angular separation for detection of a lensing system).  The
former can be understood for lensing surveys in general, whereas the
latter need to be simulated on a case-by-case basis.  

To organize our conclusions, we first give a sample application
to illustrate how our analysis may be applied to real lens surveys,
then give a more complete list of our conclusions about selection
biases in strong lensing surveys, and finally give a more general
discussion of our results.

\subsection{Example application}\label{SS:exampleapp}

As an example of how this work may be applied to the analysis
of real data, we consider the joint distribution of $\cdm$ and $\gdm$
and show how the distribution that would be observed in some strong
lensing survey is related to the intrinsic distribution.  In this
example we use the spherical higher mass model.  The concentration
is known from $N$-body simulations to have a lognormal distribution
at fixed mass \citep[e.g.,][]{2001MNRAS.321..559B}.  The distribution
of the logarithmic slope is not well known, however, because of
difficulties in resolving small scales.  Indeed, some recent work
has suggested that DM halo profiles do not converge to a single
asymptotic logarithmic slope \citep[e.g.,][]{2004MNRAS.349.1039N},
although even in that case one can imagine constraining (in both
simulations and real data) the inner slope at some fixed fraction
of the virial radius and calling that \gdm.  We assume, for
illustration, that this procedure may give a mean
$\langle\gdm\rangle=1$ and a Gaussian scatter of $0.25$.  We expect
that any constraints on \gdm\ and \cdm\ based on measuring enclosed
mass will cause \gdm\ and \cdm\ to be anti-correlated, so we assume
a correlation coefficient of $-0.3$ between \gdm\ and $\ln{(\cdm)}$.
We emphasize that the \gdm\ distribution and the correlation
coefficient are merely guesses made for the purpose of this
exercise, which means that our results should \emph{not} be taken
as a prediction for what would be seen in reality; they are intended
just to be a useful illustration.

\begin{figure*}
\begin{center}
\includegraphics[width=0.75\textwidth,angle=0]{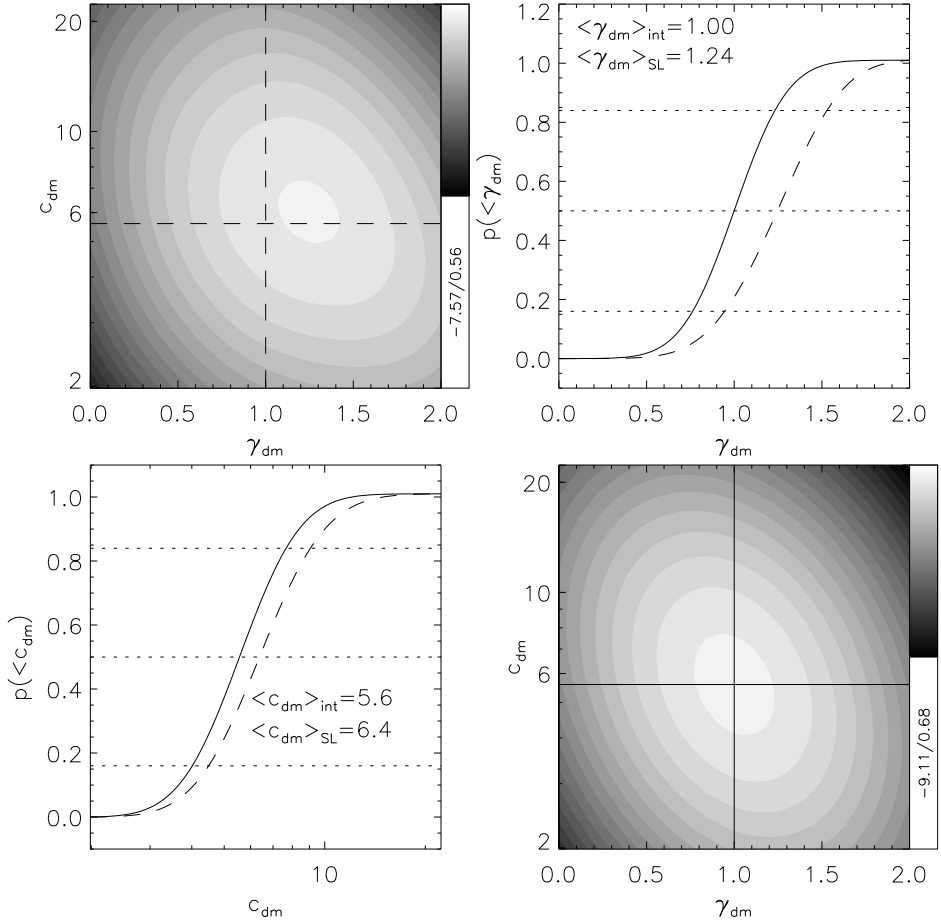} 
\caption{\label{F:inversion} 
Illustration of the procedure for removing the effects of lensing
selection bias from the observed distribution of parameters, shown
for the spherical, higher mass model.  The top left panel shows the
joint distribution $p_{\mathrm{SL}}(\gdm,\cdm)$ as it may be observed
in some strong lensing survey.  The bottom right panel shows the
corresponding intrinsic distribution $p_{\mathrm{int}}(\gdm,\cdm)$,
which would be recovered from the observed distribution by removing the
selection bias function shown in Fig.~\ref{F:gcgrid-fine}.  In both
cases the grayscale and contours are logarithmic, and the horizontal
and vertical lines are drawn at the median values of \gdm\ and \cdm\
for the \emph{intrinsic} distribution.  We emphasize that the instrinsic
joint distribution is not well known from observations or simulations,
so we have constructed a simple example for illustration (see text
for more discussion); thus, this figure should \emph{not} be taken
as a prediction for what will be seen in lensing surveys.  The top
right and bottom left panels show cumulative distributions for \gdm\
and \cdm\ (respectively) obtained by taking slices through the joint
distributions the intrinsic median values of \cdm\ or \gdm.  In each
case the dashed line represents the ``observed'' distribution while
the solid line represents the recovered intrinsic distribution.
The dotted horizontal lines indicate the median and $\pm 1\sigma$
range.
}
\end{center}
\end{figure*}

Fig.~\ref{F:inversion} shows our assumed intrinsic distribution
(lower right panel), along with the distribution that would be
observed among lens galaxies (upper right panel).  The two
distributions are related via the lensing selection function shown
in Fig.~\ref{F:gcgrid-fine}.  (For this exercise we use the selection
function without magnification bias.)  Lensing selection effects
clearly shift the peak of the joint distribution to higher values
of \gdm\ and \cdm.  To further quantify the effects, we also show
1-d slices through the joint distribution at fixed values of \gdm\
or \cdm.  These plots demonstrate that the \gdm\ and \cdm\
distributions are both shifted and broadened in the ``observed''
distribution compared with the intrinsic distribution.  In
particular, the mean and median values are shifted by $24$ (\gdm)
and $14$ (\cdm) per cent.  Consequently, any attempt to recover
the mean of \cdm\ and/or \gdm\ for some sample of strong lenses
need to be corrected for selection bias before comparing against
predictions that were made for a more general galaxy sample.

\subsection{Summary}

The biases in \gdm\ and \cdm\ are just one example of the
physical selection effects we have explored.  Here is a more complete
list of our main conclusions: 
\begin{itemize}
\item There is a strong preference for lensing systems to be at higher
  mass. This trend is counterbalanced by the strongly decreasing mass
  function at high mass, but will nonetheless significantly increase
  the mean mass for lensing systems relative to that for all galaxies.
  The degree of preference depends sensitively on how the mass is
  distributed in the inner parts of the galaxy, and on the details of
  the magnification bias (flux limit and detection strategy).
  [Section~\ref{SS:basicresults}]
\item Lensing systems show a significant preference for certain
  orientations, with factors of 2--3 variation in lensing cross-sections
  depending on their shape. In particular, the total cross-section is
  maximised when the surface mass density is maximised, i.e., when the
  galaxy is viewed with the long axis along the line of sight; and the
  four-image cross-section is maximised when the projected ellipticity
  is maximised, i.e., when the intermediate axis aligns with the line
  of sight. [Sections~\ref{SS:resultsorient} and~\ref{SS:sersicorient}]
\item The total, orientation-averaged lensing cross-section is not
  strongly dependent on the intrinsic shape of the galaxy.  In 
  contrast, the quad/double ratio may be a powerful probe of the
  distribution of the shape of the mass density in the inner parts of
  galaxies, although it also has a strong dependence on observational
  issues such as the source luminosity function.
  [Section~\ref{SS:resultsshape}]
\item An observational scarcity of three-image lens systems may place
  a strong constraint on the abundance of strongly oblate matter
  distributions. [Section~\ref{SS:resultsshape}]
\item Statistical variations in the dark matter density properties,
  such as the inner slope $\gdm$ and the concentration $\cdm$, may
  lead to a strong selection bias amounting to factors of many in
  $\st$, and to tens of per cent variation in $\sig{4}/\sig{2}$,
  nearly independent of the shape of the galaxy model and the
  observational issues relating to the magnification bias.
  Consequently, even if analyses are able to break the degeneracy
  between these parameters, for example using strong lensing plus some
  other technique to measure the profile at larger scales, they still
  cannot claim to measure the intrinsic distribution of these
  quantities directly. [Sections~\ref{SS:resultsgamma},
  \ref{SS:resultsconcgamma}, and~\ref{SS:resultsconctriax}]
\item Adiabatic contraction has a less significant effect on the
  lensing cross-section than increasing the dark matter inner slope
  for a generalised NFW model from $\gdm=1$ to $\gdm=1.5$.  However,
  the increase may still be a factor of two, which means that if
  galaxy formation scenarios dictate whether AC occurs, there may be a
  selection bias towards those galaxies that have undergone
  significant adiabatic contraction, and against those that have not.
  [Sections~\ref{SS:resultsgamma} and~\ref{SS:resultsAC}]
\item The lensing cross-sections of deprojected S\'ersic models depend
  on the indices $\ns$ of the stellar and $\ndm$ of the dark matter
  component.  Higher S\'ersic indices correspond to a more
  concentrated profile and therefore a higher lensing cross-section at
  fixed mass.  Reasonable changes of these indices, within the
  observational scatter, can change the lensing cross section by
  factors of several.  For the fiducial S\'ersic model, \Rein\ and
  \st\ are higher than for cusped models with the ``canonical'' inner
  slope value $\gdm=1$, and closer to those for the $\gdm=1.5$ cusped
  model. [Section~\ref{S:sersic}]
\item While varying the lens and source redshifts changes the range of
  physical scales probed using strong lensing, we have found that the
  dependence on $\Sigma_c$ is not very strong, and therefore that our
  conclusions regarding the extent of selection biases are valid for a
  wide range of lens-source redshift combinations
  (Fig.~\ref{F:sigmacrit}).  Furthermore, our results suggest that the
  physical selection bias is modified (when changing lens and/or
  source redshift) such that the very inner part of the profile is
  more (less) important if $\Sigma_c$ is higher (lower), making
  selection bias more (less) important than for our fiducial
  combination of lens and source redshifts. [Section~\ref{S:robustness}]
\end{itemize}

\subsection{Discussion}

Where possible, we have compared our conclusions on physical selection
biases with earlier papers in the relevant sections.  In general our
conclusions are consistent with those of previous studies, but our
pipeline allows a more unified approach to selection biases in lensing
surveys while using realistic galaxy models.  This unified
  approach allows for non-trivial extensions of previous results, to
  two-component models, different profile types, and more shapes than
  were used in previous studies. Furthermore, we can understand trends
  seen in strong lens analyses based on N-body plus semi-analytic
  modeling \citep{2007MNRAS.382..121H, 2007MNRAS.379.1195M,
    2008MNRAS.386.1845H} in terms of the interplay between different
  types of selection biases. For example, the tendency found by
  \citet{2007MNRAS.379.1195M} for quad lenses to have a pronounced
  disk component comes from an interplay between mass bias, shape
  bias, and orientation bias, as discussed in
  Section~\ref{SS:quaddouble}.  A comparison with previous results
  also teaches us lessons about the limitations of these simple
  models: for example, that when estimating selection biases with dark
  matter halo parameters, we must include a baryonic component so that
  we do not overestimate the severity of the selection bias, as
  discussed in Sections~\ref{SS:resultsgamma}
  and~\ref{SS:resultsconcgamma}.

Finally, our work has revealed a number of useful points related to
the interpretation of observational strong lensing results in general.
For example, despite the relative complexity of our mass models, the
image separation distribution in all cases has a mean near $2\Rein$
and a fairly small width (Section~\ref{SS:resultsimagesep}), as in the
isothermal case.  The dependence of this distribution on magnification
bias is a very small effect.  On a related note, these models are very
close to isothermal near \Rein\ despite strong variations at smaller
or larger scales (see  paper~I figure 5), so a measurement to
constrain the slope of the projected density profile at \Rein\ does
not allow discrimination between profiles with different \gdm\ or
\cdm, unless another technique is used to constrain the profile
independently on much smaller or larger scales (see also Paper~I).
Note, however, that because of the significant deviations from
isothermality on other scales, these models have very different
relationships between \Rein\ and \st\ than for an isothermal model
(see also Table~\ref{T:basic}).

We can apply several of our conclusions to previously-published work.
For example, \cite{2006ApJ...649..599K} use a joint lensing and
kinematics analysis of 15 SLACS lenses to infer that the average slope
of the 3d density profile near the Einstein radius is
$\langle\gamma\rangle = 2.01^{+0.02}_{-0.03}$ (68 per cent CL), with
intrinsic rms scatter of 0.12.  We defer the consideration of possible
modeling biases to future work, and consider the possible impact of
selection biases here.  This extremely high level of isothermality is,
at first glance, surprising; but then we realise from  
 paper~I figures 5 and 7 
that in fact many different combinations of
stellar and DM models can conspire to yield a total model that is
nearly isothermal over a range of scales including the Einstein
radius.  Consequently, a measurement of the total density profile in
the vicinity of the Einstein radius cannot, on its own, uniquely
determine the form of the DM profile.  However, once
\cite{2006ApJ...649..599K} impose a prior on the stellar mass-to-light
ratio they are able to constrain the projected dark matter fraction,
which (in combination with knowledge of the projected luminosity
profile) might provide constraints on the inner dark matter profile.  
However, due to the partial degeneracy between mass, \gdm\ 
and \cdm\ in strong lensing properties, this constraint does not allow
for an explicit determination of the dark matter profile parameters
without additional constraints from larger radii.  In other words,
there are many models in that three-dimensional space that would give
the observed \Rein\ and \fpdm. 


There are a number of selection bias-related problems that involve a
higher level of complexity and that we have therefore reserved for
future work. These include, but are not limited to: (i) the effect of
varying stellar mass-to-light ratios for a given galaxy model; (ii)
modeling satellite galaxies that have their own stellar and dark
matter components but are distributed within the halo of a group or
cluster (rather than residing at the centre), and may have the dark
matter and/or stars tidally stripped; (iii) lensing of extended
sources rather than point sources, for which the basic trends
  should be qualitatively similar, but for which observational
  selection effects and observing conditions are crucial; (iv) more
useful ways of quantifying image separation distributions for systems
with more than two images, and what information can be learned in this
way; (v) intrinsic misalignment of the non-spherical dark matter and
stellar components, warps, and other more complex shape models.

In addition to this future work on selection biases, we also have
designed this pipeline so that it can be easily adapted to study
modeling biases, which can occur when using overly simplified mass
models. Work on modeling biases can be extended to include strong $+$
weak lensing analyses, as can be done at cluster mass scales.
With the addition of kinematics, we can study modeling biases not just
in strong lensing systems, but also in kinematic and in kinematic $+$
strong lensing analyses. In short, this pipeline should be a useful
tool in understanding the analysis requirements for current and future
strong lensing, kinematics, and weak lensing surveys that aim to
understand galaxy density profiles.

Finally, we note that considerable additional work would be necessary
to use these results to estimate selection biases in a real survey.
For example, observational selection biases need to be accounted for.
This may include the imposition of a maximum or a minimum angular
separation between images depending on the use of an aperture or a
resolution-limited approach, respectively.  For an example of work
considering these and related issues, see \cite{1991ApJ...379..517K}.
While this may change the effects of physical selection biases, either
enhancing or counteracting them, it can also alter the lens redshift
distribution by favouring higher or lower redshifts, respectively, due
to the conversion between angular and physical scale. Furthermore, the
evolution of the survey flux limit with the source redshift
distribution must be taken into account: we have considered a fixed
\emph{luminosity} limit, but a typical flux-limited survey would have
a variable luminosity limit that gets brighter for higher redshift,
altering the biased cross-sections as a function of source redshift
even if the source luminosity function is assumed to be independent of
redshift.  This will tend to push the source galaxy population to
higher redshifts on average.   

Another point is that we have not
  attempted to model the full intrinsic distribution of galaxy
  properties, $p_{\mathrm{int}}(\vec{x})$. We have studied instead the (physical)
  mapping function $\sigma(\vec{x})$ that determines the final
  observed strong lensing population from the intrinsic one. To
  predict strong lensing populations would require a full simulation
  of the intrinsic galaxy property distribution $p_{\mathrm{int}}(\vec{x})$, which is
  clearly beyond the scope of this paper, but which has been attempted
  in limited form using N-body simulations plus semi-analytic models
  by \citet{2007MNRAS.382..121H}, \citet{2007MNRAS.379.1195M}, and
  \citet{2008MNRAS.386.1845H}.  In short, a full simulation of the
lens galaxy population and source quasar population would be necessary
to estimate the parameters of the strong lensing galaxy population in
detail for a particular survey. Nonetheless, our work is an
  important first step in unraveling and disentangling physical
  sources of selection biases for current and future strong lensing
  surveys.


\section*{Acknowledgements}
\label{S:acknowledgments}

We are thankful to Chung-Pei Ma, Michael Kuhlen and Scott Tremaine for
stimulating discussions on related topics. 
We thank the referees for constructive comments and suggestions.
This work has made use of the public \textsc{contra} software package
provided by Oleg Gnedin to perform adiabatic contraction calculations.

RM and GvdV acknowledge support provided by NASA through Hubble
Fellowship grants HST-HF-01199.02-A and HST-HF-01202.01-A
(respectively), awarded by the Space Telescope Science Institute,
which is operated by the Association of Universities for Research in
Astronomy, Inc., for NASA, under contract NAS 5-26555.
CRK acknowledges support from NSF through grant AST-0747311, and from
NASA through grant HST-AR-11270.01-A from the Space Telescope Science
Institute, which is operated by the Association of Universities for
Research in Astronomy, Inc., under NASA contract NAS 5-26555.


\bibliographystyle{mn2e}


\begin{thebibliography}{61}
\expandafter\ifx\csname natexlab\endcsname\relax\def\natexlab#1{#1}\fi

\bibitem[{{Baltz} {et~al.}(2009){Baltz}, {Marshall}, \&
  {Oguri}}]{2009JCAP...01..015B}
{Baltz} E.~A., {Marshall} P., {Oguri} M., 2009, JCAP, 1, 15

\bibitem[{{Blumenthal} {et~al.}(1986){Blumenthal}, {Faber}, {Flores}, \&
  {Primack}}]{1986ApJ...301...27B}
{Blumenthal} G.~R., {Faber} S.~M., {Flores} R., {Primack} J.~R., 1986, \apj,
  301, 27

\bibitem[{{Bolton} {et~al.}(2008){Bolton}, {Burles}, {Koopmans}, {Treu},
  {Gavazzi}, {Moustakas}, {Wayth}, \& {Schlegel}}]{2008ApJ...682..964B}
{Bolton} A.~S., {Burles} S., {Koopmans} L.~V.~E., {Treu} T., {Gavazzi} R.,
  {Moustakas} L.~A., {Wayth} R., {Schlegel} D.~J., 2008, \apj, 682, 964

\bibitem[{{Browne} {et~al.}(2003){Browne}, {Wilkinson}, {Jackson}, {Myers},
  {Fassnacht}, {Koopmans}, {Marlow}, {Norbury}, {Rusin}, {Sykes}, {Biggs},
  {Blandford}, {de Bruyn}, {Chae}, {Helbig}, {King}, {McKean}, {Pearson},
  {Phillips}, {Readhead}, {Xanthopoulos}, \& {York}}]{2003MNRAS.341...13B}
{Browne} I.~W.~A. et~al. 2003, \mnras, 341, 13

\bibitem[{{Bullock} {et~al.}(2001){Bullock}, {Kolatt}, {Sigad}, {Somerville},
  {Kravtsov}, {Klypin}, {Primack}, \& {Dekel}}]{2001MNRAS.321..559B}
{Bullock} J.~S., {Kolatt} T.~S., {Sigad} Y., {Somerville} R.~S., {Kravtsov}
  A.~V., {Klypin} A.~A., {Primack} J.~R., {Dekel} A., 2001, \mnras, 321, 559

\bibitem[{{Cardone}(2004)}]{2004A&A...415..839C}
{Cardone} V.~F., 2004, \aap, 415, 839

\bibitem[{{Chae}(2003)}]{2003MNRAS.346..746C}
{Chae} K.-H., 2003, \mnras, 346, 746

\bibitem[{{Ciotti} \& {Bertin}(1999)}]{1999A&A...352..447C}
{Ciotti} L., {Bertin} G., 1999, \aap, 352, 447

\bibitem[{{Comerford} \& {Natarajan}(2007)}]{2007MNRAS.379..190C}
{Comerford} J.~M., {Natarajan} P., 2007, \mnras, 379, 190

\bibitem[{{Dobler} {et~al.}(2008){Dobler}, {Keeton}, {Bolton}, \&
  {Burles}}]{2008ApJ...685...57D}
{Dobler} G., {Keeton} C.~R., {Bolton} A.~S., {Burles} S., 2008,
\apj, 685, 57

\bibitem[{{Falco} {et~al.}(2001){Falco}, {Kochanek}, {Leh{\'a}r}, {McLeod},
  {Mu{\~n}oz}, {Impey}, {Keeton}, {Peng}, \& {Rix}}]{2001ASPC..237...25F}
{Falco} E.~E. et~al, 2001, in
  Astronomical Society of the Pacific Conference Series, Vol. 237,
  {Brainerd} T.~G.,
  {Kochanek} C.~S., eds., 25

\bibitem[{{Fassnacht} {et~al.}(2004){Fassnacht}, {Marshall}, {Baltz},
  {Blandford}, {Schechter}, \& {Tyson}}]{2004AAS...20510827F}
{Fassnacht} C.~D., {Marshall} P.~J., {Baltz} A.~E., {Blandford} R.~D.,
  {Schechter} P.~L., {Tyson} J.~A., 2004, in Bulletin of the American
  Astronomical Society, 36, 1531

\bibitem[{{Fedeli} {et~al.}(2007){Fedeli}, {Bartelmann}, {Meneghetti}, \&
  {Moscardini}}]{2007A&A...473..715F}
{Fedeli} C., {Bartelmann} M., {Meneghetti} M., {Moscardini} L., 2007, \aap,
  473, 715

\bibitem[{{Ferrarese} {et~al.}(2006){Ferrarese}, {C{\^o}t{\'e}}, {Jord{\'a}n},
  {Peng}, {Blakeslee}, {Piatek}, {Mei}, {Merritt}, {Milosavljevi{\'c}},
  {Tonry}, \& {West}}]{2006ApJS..164..334F}
{Ferrarese} L. et~al. 2006, \apjs, 164, 334

\bibitem[{{Fukugita} \& {Turner}(1991)}]{1991MNRAS.253...99F}
{Fukugita} M., {Turner} E.~L., 1991, \mnras, 253, 99

\bibitem[{{Gnedin} {et~al.}(2004){Gnedin}, {Kravtsov}, {Klypin}, \&
  {Nagai}}]{2004ApJ...616...16G}
{Gnedin} O.~Y., {Kravtsov} A.~V., {Klypin} A.~A., {Nagai} D., 2004, \apj, 616,
  16

\bibitem[{{Hernquist}(1990)}]{1990ApJ...356..359H}
{Hernquist} L., 1990, \apj, 356, 359

\bibitem[{{Hilbert} {et~al.}(2007){Hilbert}, {White}, {Hartlap}, \&
    {Schneider}}]{2007MNRAS.382..121H} {Hilbert} S., {White} S.~D.~M.,
  {Hartlap} J., {Schneider} P., 2007, \mnras, 382, 121 

\bibitem[{{Hilbert} {et~al.}(2008){Hilbert}, {White}, {Hartlap}, \&
    {Schneider}}]{2008MNRAS.386.1845H} {Hilbert} S., {White} S.~D.~M.,
  {Hartlap} J., {Schneider} P., 2008, \mnras, 386, 1845 

\bibitem[{{Huterer} {et~al.}(2005){Huterer}, {Keeton}, \&
  {Ma}}]{2005ApJ...624...34H}
{Huterer} D., {Keeton} C.~R., {Ma} C.-P., 2005, \apj, 624, 34

\bibitem[{{Huterer} \& {Ma}(2004)}]{2004ApJ...600L...7H}
{Huterer} D., {Ma} C.-P., 2004, \apjl, 600, L7

\bibitem[{{Inada} {et~al.}(2008){Inada}, {Oguri}, {Becker}, {Shin}, {Richards},
  {Hennawi}, {White}, {Pindor}, {Strauss}, {Kochanek}, {Johnston}, {Gregg},
  {Kayo}, {Eisenstein}, {Hall}, {Castander}, {Clocchiatti}, {Anderson},
  {Schneider}, {York}, {Lupton}, {Chiu}, {Kawano}, {Scranton}, {Frieman},
  {Keeton}, {Morokuma}, {Rix}, {Turner}, {Burles}, {Brunner}, {Sheldon},
  {Bahcall}, \& {Masataka}}]{2008AJ....135..496I}
{Inada} N. et~al. 2008, \aj, 135, 496

\bibitem[{{Jiang} \& {Kochanek}(2007)}]{2007ApJ...671.1568J} {Jiang}
  G., {Kochanek} C.~S., 2007, \apj, 671, 1568

\bibitem[{{Kauffmann} {et~al.}(2003){Kauffmann}, {Heckman}, {White}, {Charlot},
  {Tremonti}, {Brinchmann}, {Bruzual}, {Peng}, {Seibert}, {Bernardi},
  {Blanton}, {Brinkmann}, {Castander}, {Cs{\'a}bai}, {Fukugita}, {Ivezic},
  {Munn}, {Nichol}, {Padmanabhan}, {Thakar}, {Weinberg}, \&
  {York}}]{2003MNRAS.341...33K}
{Kauffmann} G. et~al. 2003, \mnras, 341, 33

\bibitem[{{Keeton}(2001{\natexlab{a}})}]{2001ApJ...561...46K}
{Keeton} C.~R., 2001{\natexlab{a}}, \apj, 561, 46

\bibitem[{{Keeton}(2001{\natexlab{b}})}]{2001astro.ph..2340K}
---, 2001{\natexlab{b}}, preprint (astro-ph/0102340)

\bibitem[{{Keeton} \& {Kochanek}(1998)}]{1998ApJ...495..157K}
{Keeton} C.~R., {Kochanek} C.~S., 1998, \apj, 495, 157

\bibitem[{{Keeton} {et~al.}(1997){Keeton}, {Kochanek}, \&
  {Seljak}}]{1997ApJ...482..604K}
{Keeton} C.~R., {Kochanek} C.~S., {Seljak} U., 1997, \apj, 482, 604

\bibitem[{{Keeton} \& {Madau}(2001)}]{2001ApJ...549L..25K}
{Keeton} C.~R., {Madau} P., 2001, \apjl, 549, L25

\bibitem[{{Keeton} \& {Zabludoff}(2004)}]{2004ApJ...612..660K}
{Keeton} C.~R., {Zabludoff} A.~I., 2004, \apj, 612, 660

\bibitem[{{Keeton}(1998)}]{1998PhDT.........6K}
{Keeton} C.~R.~I., 1998, PhD thesis

\bibitem[{{Kochanek}(1991)}]{1991ApJ...379..517K}
{Kochanek} C.~S., 1991, \apj, 379, 517

\bibitem[{{Kochanek}(1996)}]{1996ApJ...473..595K}
---, 1996, \apj, 473, 595

\bibitem[{{Kochanek}(2006)}]{Saas-Fee}
---, 2006, in Saas-Fee Advanced Course 33: Gravitational Lensing: Strong, Weak
  and Micro, {Meylan} G., {Jetzer} P., {North} P., {Schneider} P., {Kochanek}
  C.~S., {Wambsganss} J., eds., 91

\bibitem[{{Kochanek} \& {White}(2001)}]{2001ApJ...559..531K}
{Kochanek} C.~S., {White} M., 2001, \apj, 559, 531

\bibitem[{{Koopmans} {et~al.}(2004){Koopmans}, {Browne}, \&
  {Jackson}}]{2004NewAR..48.1085K}
{Koopmans} L.~V.~E., {Browne} I.~W.~A., {Jackson} N.~J., 2004, New Astronomy
  Review, 48, 1085

\bibitem[{{Koopmans} {et~al.}(2006){Koopmans}, {Treu}, {Bolton}, {Burles}, \&
  {Moustakas}}]{2006ApJ...649..599K}
{Koopmans} L.~V.~E., {Treu} T., {Bolton} A.~S., {Burles} S., {Moustakas} L.~A.,
  2006, \apj, 649, 599

\bibitem[{{Kuhlen} {et~al.}(2004){Kuhlen}, {Keeton}, \&
  {Madau}}]{2004ApJ...601..104K}
{Kuhlen} M., {Keeton} C.~R., {Madau} P., 2004, \apj, 601, 104

\bibitem[{{Li} \& {Ostriker}(2002)}]{2002ApJ...566..652L}
{Li} L.-X., {Ostriker} J.~P., 2002, \apj, 566, 652

\bibitem[{{Maller} {et~al.}(1997){Maller}, {Flores}, \&
  {Primack}}]{1997ApJ...486..681M}
{Maller} A.~H., {Flores} R.~A., {Primack} J.~R., 1997, \apj, 486, 681

\bibitem[{{Mandelbaum} {et~al.}(2006){Mandelbaum}, {Seljak}, {Kauffmann},
  {Hirata}, \& {Brinkmann}}]{2006MNRAS.368..715M}
{Mandelbaum} R., {Seljak} U., {Kauffmann} G., {Hirata} C.~M., {Brinkmann} J.,
  2006, \mnras, 368, 715

\bibitem[{{Marshall} {et~al.}(2005){Marshall}, {Blandford}, \&
  {Sako}}]{2005NewAR..49..387M}
{Marshall} P., {Blandford} R., {Sako} M., 2005, New Astronomy Review, 49, 387

\bibitem[{{Merritt} {et~al.}(2005){Merritt}, {Navarro}, {Ludlow}, \&
  {Jenkins}}]{2005ApJ...624L..85M}
{Merritt} D., {Navarro} J.~F., {Ludlow} A., {Jenkins} A., 2005, \apjl, 624, L85

\bibitem[{{M{\"o}ller} {et~al.}(2007){M{\"o}ller}, {Kitzbichler}, \&
  {Natarajan}}]{2007MNRAS.379.1195M}
{M{\"o}ller} O., {Kitzbichler} M., {Natarajan} P., 2007, \mnras, 379, 1195

\bibitem[{{Momcheva} {et~al.}(2006){Momcheva}, {Williams}, {Keeton}, \&
  {Zabludoff}}]{2006ApJ...641..169M}
{Momcheva} I., {Williams} K., {Keeton} C., {Zabludoff} A., 2006, \apj, 641, 169

\bibitem[{{Morganti} {et~al.}(2006){Morganti}, {de Zeeuw}, {Oosterloo},
  {McDermid}, {Krajnovi{\'c}}, {Cappellari}, {Kenn}, {Weijmans}, \&
  {Sarzi}}]{2006MNRAS.371..157M}
{Morganti} R. et~al. 
  2006, \mnras, 371, 157

\bibitem[{{Navarro} {et~al.}(1997){Navarro}, {Frenk}, \&
  {White}}]{1997ApJ...490..493N}
{Navarro} J.~F., {Frenk} C.~S., {White} S.~D.~M., 1997, \apj, 490, 493

\bibitem[{{Navarro} {et~al.}(2004){Navarro} et~al.}]{2004MNRAS.349.1039N} {Navarro} J.~F. et~al. 2004, \mnras, 349, 1039

\bibitem[{{Oguri}(2007)}]{2007NJPh....9..442O}
{Oguri} M., 2007, New Journal of Physics, 9, 442

\bibitem[{{Oguri} \& {Keeton}(2004)}]{2004ApJ...610..663O}
{Oguri} M., {Keeton} C.~R., 2004, \apj, 610, 663

\bibitem[{{Oguri} {et~al.}(2002){Oguri}, {Taruya}, {Suto}, \&
  {Turner}}]{2002ApJ...568..488O}
{Oguri} M., {Taruya} A., {Suto} Y., {Turner} E.~L., 2002, \apj, 568, 488

\bibitem[{{Porciani} \& {Madau}(2000)}]{2000ApJ...532..679P}
{Porciani} C., {Madau} P., 2000, \apj, 532, 679

\bibitem[{{Prugniel} \& {Simien}(1997)}]{1997A&A...321..111P}
{Prugniel} P., {Simien} F., 1997, \aap, 321, 111

\bibitem[{{Puchwein} {et~al.}(2005){Puchwein}, {Bartelmann}, {Dolag}, \&
  {Meneghetti}}]{2005A&A...442..405P}
{Puchwein} E., {Bartelmann} M., {Dolag} K., {Meneghetti} M., 2005, \aap, 442,
  405

\bibitem[{{Read} \& {Trentham}(2005)}]{2005RSPTA.363.2693R}
{Read} J.~I., {Trentham} N., 2005, Royal Society of London Philosophical
  Transactions Series A, 363, 2693

\bibitem[{{Rozo} {et~al.}(2007){Rozo}, {Chen}, \&
  {Zentner}}]{2007arXiv0710.1683R}
{Rozo} E., {Chen} J., {Zentner} A.~R., 2007, preprint (arXiv:0710.1683)

\bibitem[{{Rozo} {et~al.}(2006){Rozo}, {Nagai}, {Keeton}, \&
  {Kravtsov}}]{2008ApJ...687...22R}
{Rozo} E., {Nagai} D., {Keeton} C., {Kravtsov} A., 2008, \apj, 687, 22

\bibitem[{{Rusin} \& {Tegmark}(2001)}]{2001ApJ...553..709R}
{Rusin} D., {Tegmark} M., 2001, \apj, 553, 709

\bibitem[{{Sellwood} \& {McGaugh}(2005)}]{2005ApJ...634...70S}
{Sellwood} J.~A., {McGaugh} S.~S., 2005, \apj, 634, 70

\bibitem[{{S\'ersic}(1968)}]{1968adga.book.....S}
{S\'ersic} J.~L., 1968, {Atlas de galaxias australes}. Cordoba, Argentina:
  Observatorio Astronomico, 1968

\bibitem[{{Turner} {et~al.}(1984){Turner}, {Ostriker}, \&
  {Gott}}]{1984ApJ...284....1T}
{Turner} E.~L., {Ostriker} J.~P., {Gott} III J.~R., 1984, \apj, 284, 1

\bibitem[{{Wambsganss} {et~al.}(2008){Wambsganss}, {Ostriker}, \&
  {Bode}}]{2008ApJ...676..753W}
{Wambsganss} J., {Ostriker} J.~P., {Bode} P., 2008, \apj, 676, 753

\bibitem[{{Wyithe} {et~al.}(2001){Wyithe}, {Turner}, \&
  {Spergel}}]{2001ApJ...555..504W}
{Wyithe} J.~S.~B., {Turner} E.~L., {Spergel} D.~N., 2001, \apj, 555, 504

\bibitem[{{Young}(1980)}]{1980ApJ...242.1232Y}
{Young} P., 1980, \apj, 242, 1232

\end{thebibliography}


\appendix

\section{Variation of lensing cross-section with inner
  slope and orientation}
\label{A:gammaorient}

In Section~\ref{SS:resultsgamma}, we quantified a selection bias based on
dark matter halo inner slope using ratios
\begin{equation}
  \label{eq:ratiosigN}
  \langle R_{1,0.5}(N)\rangle=\frac{\langle
    \sig{N}(\gdm=1.0)\rangle}{\langle\sig{N}(\gdm=0.5)\rangle},
\end{equation}
where the averages indicate an average over orientations, and $N$ can
indicate either the total cross-section or the cross-section for some
number of images $N$. We can also define a comparable quantity
($\langle R_{1.5,1}(N)\rangle$) for $\gdm=1.5$ versus $\gdm=1$.
Ratios of one indicate an absence of selection bias, and we have
generally found these ratios to be $>1$ (the strong-lensing
cross-sections increase with $\gdm$).

We now address the question of whether these ratios are the same for
each viewing direction, i.e. we consider
\begin{equation}
  \label{eq:ratiosigNorient}
  R_{1,0.5}(N,a'/a,b'/a) =
  \frac{\sig{N}(\gdm=1.0,a'/a,b'/a)}{\sig{N}(\gdm=0.5,a'/a,b'/a)}
\end{equation}
and a comparable quantity $R_{1.5,1}(N,a'/a,b'/a)$. If these ratios stay
(nearly) constant while changing $a'/a$ and $b'/a$, then the selection
bias with dark matter halo inner slope is not a (strong) function of
viewing direction for galaxies with non-spherical shapes.

For the triaxial lower mass models, these ratios are indeed
approximately constant (to within $\sim 10$ per cent) for $\st$, $\sig{2}$,
and $\sig{4}$, independent of magnification bias mode. For the triaxial higher
mass model, where dark matter is more important, we find a
stronger dependence on viewing direction. For $\sig{2}$, there are up
to $\sim 30$ per cent level trends for $R_{1,0.5}$ and $R_{1.5,1}$ to be
lower (i.e., closer to unity and hence indicating less selection bias
with $\gdm$), when viewing the model along the intermediate axis.
This change occurs because in that orientation, the projected
ellipticity and therefore $\sigma_4$ is maximised, which decreases
$\sigma_2$ because $2$-image systems can only occur outside of the
tangential  caustic. 
For $\sig{4}$, this is the case when viewing the model along the short
axis (for $R_{1,0.5}$) and along the long axis (for $R_{1.5,1}$).


It is not clear that the
details of this orientation-dependence of the selection bias with
$\gdm$ is very important except for  special situations. When
analysing data from a typical survey, we expect that galaxies are
oriented randomly (before the imposition of strong lensing
``orientation bias''), so the angle-averaged selection biases from
Section~\ref{SS:resultsgamma} are the relevant quantities. We include
the analysis here as an indication that if one isolates a special
subset of strong lenses based on their projected shape, such as  those
for which the lens is particularly round or flattened, then the
treatment of selection bias with $\gdm$ and possibly other properties
must be specialised to that subset rather than using the
orientation-averaged case.

\section{Relationship between dark matter fraction and Einstein
  radius}
\label{A:dmfrac}

We seek to understand why the relation between dark matter fraction
and Einstein radius shown in Fig.~\ref{F:fdm} is perfectly one-to-one.
We begin with the Einstein radius definition, 
${\bar\kappa}(\Rein) = 1$, meaning that the average convergence
within the Einstein radius is unity (see also section 3.1 of paper I).
Suppose the model has stellar 
and dark matter components.  Then the definition of the Einstein
radius is
\begin{equation} \label{eq:dmfrac1}
  {\bar\kappa}_{\rm tot}(\Rein) =
  {\bar\kappa}_\star(\Rein) + {\bar\kappa}_{\rm dm}(\Rein) = 1\,.
\end{equation}
The stellar and dark matter mass fractions within the Einstein radius
are, by definition,
\begin{equation}
\fps(\Rein) = \frac{{\bar\kappa}_\star(\Rein)}
  {{\bar\kappa}_{\rm tot}(\Rein)}\ ,
  \quad
\fpdm(\Rein) = \frac{{\bar\kappa}_{\rm dm}(\Rein)}
  {{\bar\kappa}_{\rm tot}(\Rein)}\ .
\end{equation}
Thus, equation~(\ref{eq:dmfrac1}) can be rewritten as
\begin{equation} \label{eq:dmfrac2}
  \fps(\Rein) + \fpdm(\Rein) = 1
\end{equation}
On the one hand, this is entirely unsurprising: if the galaxy has
stars and dark matter, the stellar and dark matter mass fractions 
must add up to unity.  But on the other hand, this equation does
illuminate our results in Fig.~\ref{F:fdm}.  Consider changing the
dark matter component while keeping the stellar component fixed.
As we vary the dark matter component, the Einstein radius will
change.  That in turn changes the first term in equation~(\ref{eq:dmfrac2}),
which necessarily changes the second term, too.  However, if we
hold the stellar component fixed, the first term is a function of
a single variable, namely the Einstein radius.  Since
$\fpdm= 1 - \fps$ that means the dark matter fraction can
likewise depend only on a single variable, $\Rein$.  In other words,
if we hold the stellar component fixed, then no matter what we do
to the dark matter component, the changes to \fpdm\ propagate
only through the Einstein radius, and we have a perfect one-to-one
correspondence between $\Rein$ and \fpdm.

\end{document}